# DART ejecta reveal the recoil momentum budget of a kinetic impact

Zhijun Song[1,2], Yang Yu[1]*, Jian-Yang Li[2]*, Fabio Ferrari[3], Patrick Michel[4,5], Adriano C. Bagatin[6], Gonzalo Tancredi[7], Lioudmila Kolokolova[8], Tony F. Farnham[8], Andrew S. Rivkin[9], Bin Cheng[10], Hexi Baoyin[10]

**Affiliations:**

[1] School of Aeronautic Science and Engineering, Beihang University; Beijing 100191, China.

[2] School of Atmospheric Sciences, Sun Yat-sen University; Zhuhai, 519082, China.

[3] Department of Aerospace Science and Technology, Politecnico di Milano; Via La Masa 34, 20156 Milan, Italy.

[4] Université Côte d'Azur, Observatoire de la Côte d'Azur, CNRS, Laboratoire Lagrange; Nice, France.

[5] Department of Systems Innovation, School of Engineering, University of Tokyo; Tokyo, Japan.

[6] IUFACyT—DFISTS. Universidad de Alicante; Alicante, Spain.

[7] Departamento de Astronomía, Facultad de Ciencias; Udelar, Uruguay.

[8] Department of Astronomy, University of Maryland; College Park, MD, USA.

[9] Applied Physics Laboratory, Johns Hopkins University; Laurel, MD, USA.

[10] School of Aerospace Engineering, Tsinghua University; Beijing 100084, China.

*Corresponding author. Email: yuyang.thu@icloud.com; lijianyang@mail.sysu.edu.cn

## Abstract

Double Asteroid Redirection Test (DART) demonstrated that ejecta recoil enhances asteroid deflection, but the ejecta processes generating and distributing this recoil remained unclear. We reconstruct three-dimensional ejecta evolution from time-resolved morphology and brightness observations, recovering previously missing constraints on momentum transfer and material transport. Integrating reconstructed momentum yields momentum enhancement factors $\beta_{\mathrm{D}} = 2.13 \pm 0.19$ (1σ) for Dimorphos and $\beta_{\mathrm{sys}} = 2.04 \pm 0.19$ (1σ) for the Didymos system. Combined with the measured deflection, the reconstructed recoil constrains Dimorphos's bulk density to $1435.18^{+112.59}_{-153.71}$ kg m$^{-3}$ without assuming its mass. By linking observed orbital deflection to the released material, this reconstruction provides ejecta-based constraints for assessing kinetic-impact effectiveness and investigating how natural collisions drive mass loss and debris transport among small Solar System bodies.

## Main Text

NASA's Double Asteroid Redirection Test (DART) (*1*) demonstrated that material ejected during an asteroid collision can substantially enhance the momentum transferred to the target (*2*–*4*). However, orbital measurements reveal only how an asteroid's motion changes after an impact (*2*), but not how the ejected material contributed to that change. Different ejecta mass and velocity distributions can generate similar net recoil, so the measured deflection does not uniquely determine the underlying material ejection. Time-resolved observations of the ejecta provide further constraints on the mass and velocity distributions of the ejected material (*5*), as their evolving structures reflect both the initial ejection conditions and subsequent particle transport. Linking these observations through a dynamical reconstruction can reveal how different ejecta populations contribute to the visible aftermath of a collision, the initial recoil, and the subsequent redistribution of momentum within the asteroid system. Understanding these processes can clarify how collisions among small Solar System bodies redistribute mass and momentum, alter target motion, and shape ejecta structures. Such understanding also can elucidate the physical mechanisms of kinetic-impact deflection and inform the design and assessment of future planetary defense missions.

LICIACube (*6*) imaged the ejecta within minutes after impact (*7*, *8*), and the Hubble Space Telescope (HST) followed the plume from the first post-impact orbit through the following weeks (*9*). These observations show an asymmetric ejecta cone (*8*, *10*), filamentary and curved streams (*9*), a bifurcated tail (*9*, *11*, *12*), boulder clusters (*8*, *10*, *13*), and a temporary pause in the fading brightness (*9*, *12*). Here, we use a closed-loop three-dimensional dynamical model (See Supplementary Materials) jointly constrained by these observations to reconstruct the ejecta momentum distribution and its subsequent evolution (fig. S1). The reconstructed particle population reproduces the all observed morphology and brightness evolution, including the formation of the bifurcated tail and the temporary brightness plateau. The reconstruction reveals that fine particles dominate the visible structures, whereas larger, faster ejecta carry most of the recoil momentum. Tracking their subsequent evolution further quantifies how gravitational interactions, radiation pressure, and re-impacts affect ejecta momentum and mass transport and the momentum budget of the binary system. Integrating the reconstructed ejecta momentum constrains the momentum enhancement factors for Dimorphos, $\beta_{\mathrm{D}}$ (*14*), and for the heliocentric response of the Didymos system, $\beta_{\mathrm{sys}}$ (*15*). Combining the reconstructed recoil with orbital measurements further constrains the bulk density of Dimorphos without prescribing its mass, providing a prediction that can be independently tested by Hera.

## Morphology constrains the recoil-producing ejecta field

The early ejecta morphology fixes the launch field that produced Dimorphos recoil. As a typical representative of these physically acceptable solutions (table. S2), Fig. 1 shows direct comparisons between the simulated images from the result (Simulation set ID 18 in table S2) and the observations from LICIACube (*8*, *10*) and HST (*9*). The converged solution reproduces nearly all observed features of the DART ejecta, including the tail brightness profile (fig. S2), the filamentary streams and the curvature of the ejecta cone edge, the curvature of the ejecta streams, the secondary tail, the dark ring in the ejecta cone (*7*, *10*, *16*, *17*) and a slight pause in the decaying brightness curve within 50 km photometric aperture(*9*, *11*, *12*) (Fig. 4B). The wider cone opening angle and the curved cone edge (*11*, *12*, *20*) require a broad ejecta curtain released over a curved surface rather than a simple axisymmetric plume (*18*). The same launch field with the filamentary structures evolves into the HST streamers and tails when propagated under Didymos-Dimorphos gravity and solar radiation pressure (Fig. 1). This agreement links the early images to the later dynamical evolution and constrains the ejecta velocities that dominate the recoil (Fig. 1).

Forward modeling of the tail brightness profiles (see Method), including size-dependent filtering by the binary environment, constrains the source size-frequency distribution (fig. S2). In the representative solution, the inferred distribution follows four power-law segments below a particle radius of 1 m, with a Weibull distribution for larger boulders (table S3). The additional segment captures a flatter source distribution at radii of $3 \times 10^{-5} - 2 \times 10^{-3}$ m than that previously inferred for the tail (*9*, *19*), where larger particles are preferentially excluded from the observed tail (Fig. 2C). This size-selection effect that significantly affect the plume brightness and the momentum distribution of ejecta will be examined in detail below.

The reconstructed ejecta momentum gives $\beta_D$ = 2.13 ± 0.19 (1σ) for Dimorphos (Fig 5A). This value comes from the recoil-producing ejecta field inferred from the images, not from prescribing an assumed Dimorphos mass. The observed orbital deflection (*2*) is used subsequently to infer Dimorphos mass and density by momentum balance (See Method). This ordering separates the image-derived ejecta recoil from the density-dependent interpretation of the target response.

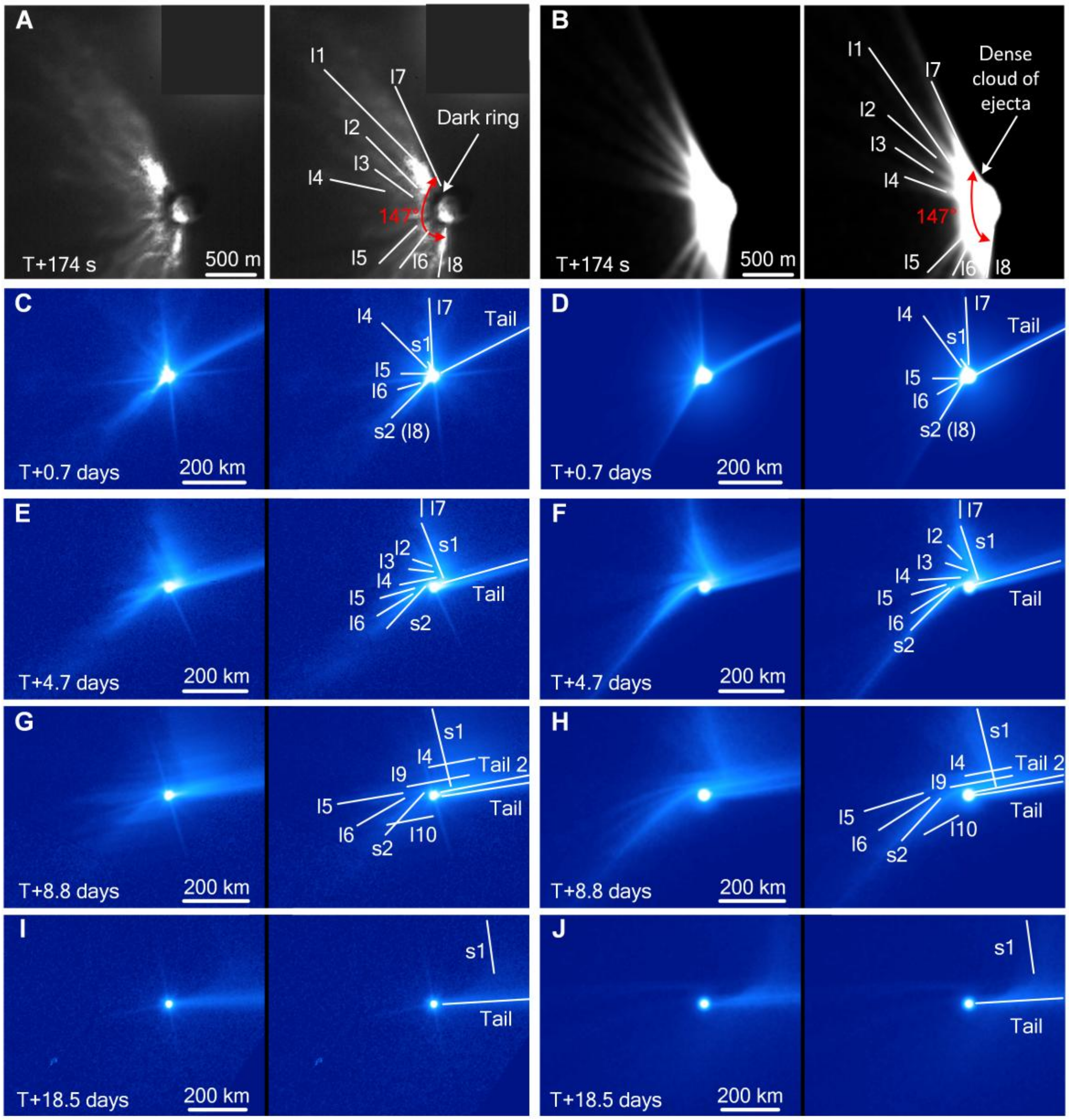

**Fig. 1. Common ejecta launch field links the early curtain to the evolving dust tails.** Observed images (**A**, **C**, **E**, **G**, and **I**) and corresponding synthetic images of the representative converged solution (**B**, **D**, **F**, **H**, and **J**) show the ejecta at 174 s and 0.7, 4.7, 8.8, and 18.5 days after the DART impact, respectively. Each panel pairs an unannotated image (left) with an annotated copy (right). (**A**) The image of ejecta cone acquired by LUKE onboard LICIACube at a distance of 71 km, 6.5 s after closest approach (*11*, *12*) (174 s after impact). It is adapted from the calibrated red filter image available in the Planetary Data System (PDS). (**B**) Synthetic images from numerical simulation with the same acquisition time, viewing geometry, and illumination conditions as the LUKE image. The LICIACube image (**A**) and synthetic image (**B**) show a broad, curved

ejecta curtain, respectively. Labels identify linear features, the observed dark ring (*12*), and the modeled dense ejecta cloud. The synthetic LICIACube image shows a dense cloud of ejecta above the impact site (**B**). This optically thick cloud (Fig. S3) casts a shadow on the ejecta material beneath, producing the observed "dark ring" that separates Dimorphos' limb from the ejecta plume (**A**). (**C** to **J**) Hubble Space Telescope images and corresponding synthetic images trace the evolution of linear features, curved streams (s1 and s2), and the main tail, including a secondary tail at 8.8 days. Matching labels indicate proposed morphological correspondences between observations and simulations. A comparison between HST observations and synthetic images of the ejecta tail over a wider field of view is shown in Fig. S4. "T" denotes the impact time. Scale bars, 500 m (**A** and **B**) and 200 km (**C** to **J**).

## Momentum generation is distributed unevenly across ejecta sizes

The reconstructed launch field separates the particles that dominate the optical signal from those that dominate the momentum budget. In the representative converged solution, particles with radii below 3 cm provide 96% of the geometric scattering cross-section and 47% of the ejected mass, and contribute 59% of $\Delta\beta_D$. The remaining particles contain 53% of the ejected mass and provide only 4% of the geometric scattering cross-section, yet contribute 41% of $\Delta\beta_D$ (Fig. 2A). This result indicates the ejecta that make the plume bright are not necessarily the ejecta that make the impact effective.

The reconstructed launch field derived enhancement is $\beta_D = 2.30$ for this representative solution. The large-particle population makes a distinct contribution to the Dimorphos launch budget. Particles with radii greater than 0.1 m carry a net momentum of approximately (-0.2, -1.0, 0.1) × $10^6$ kg m $s^{-1}$, expressed on the Dimorphos body reference frame. Their along-track momentum is approximately consistent with the estimate value derived from LICIACube observations (*10*). Differences in the transverse components may reflect a direction dependence in observations, i.e., undetected boulders could carry opposing transverse momentum and thereby reduce the net transverse components (*10*).

The joint size–speed budget further reveals the recoil momentum is dominated by a high-speed ejecta subpopulation with a very small mass fraction (Fig. 2B). Ejecta launched faster than 1 m $s^{-1}$ comprise only 4 % of the ejected mass but supply 71% of $\Delta\beta_D$, distributed across several size classes (Fig. 2B). Conversely, particles launched below 0.09 m $s^{-1}$ provide 44% of the cross-section but only 4% of the enhancement. Recoil therefore depends on the joint distribution of mass, speed and direction, rather than on optical prominence or mass alone. By linking brightness, trajectories, particle sizes, and velocities within a unified dynamical framework, ejecta modeling reveals the underlying momentum distribution that cannot be recovered from plume morphology or orbital deflection measurements alone.

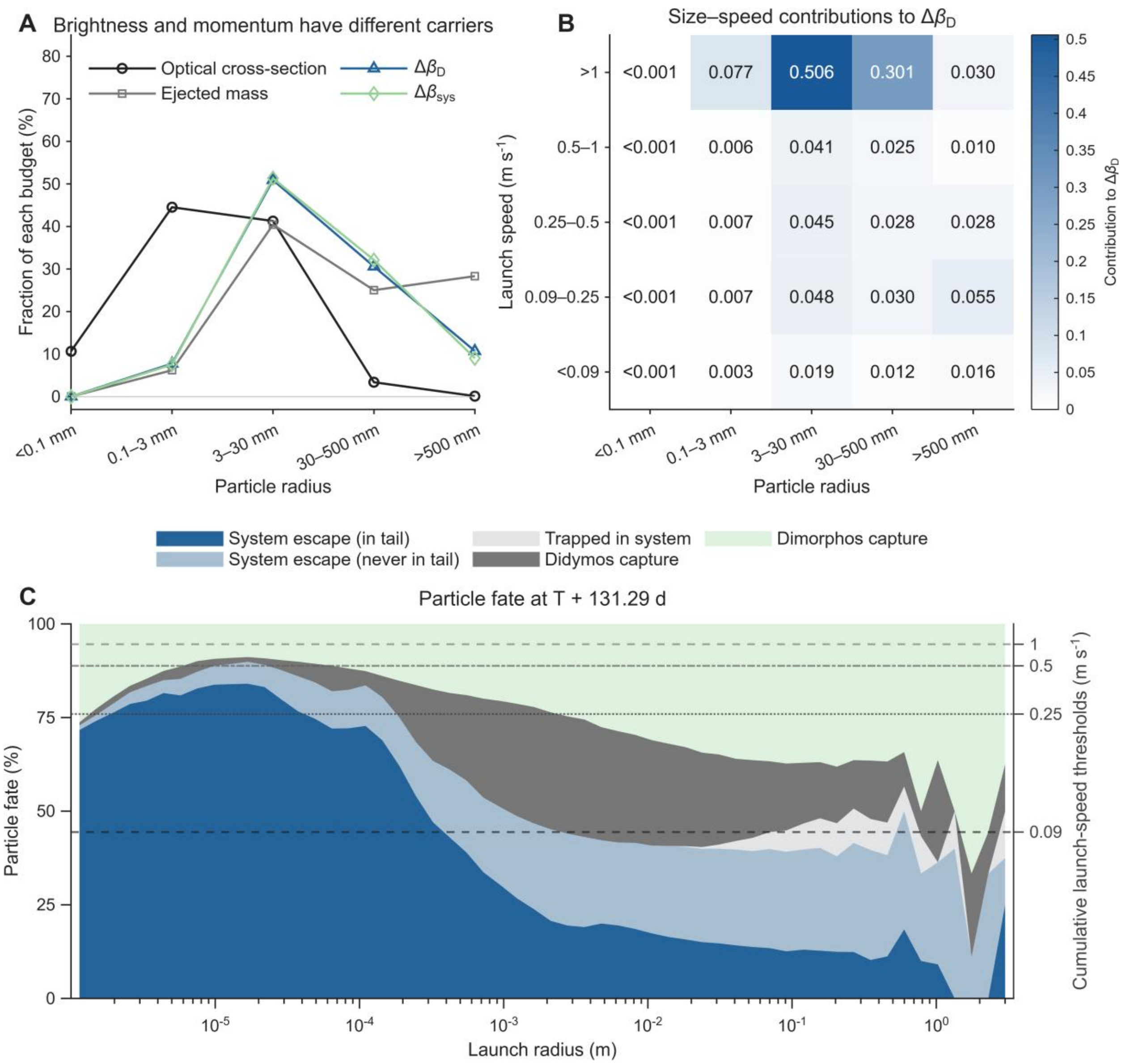

**Fig. 2**. **Size, velocity, and dynamical filtering partition the optical and recoil-momentum carriers among DART ejecta**. (**A**) Fractions of geometric scattering cross-section, ejected material mass, the Dimorphos momentum enhancement ($\Delta\beta_{\mathrm{D}} = \beta_{\mathrm{D}} - 1$), and system heliocentric momentum enhancement ($\Delta\beta_{\mathrm{sys}} = \beta_{\mathrm{sys}} - 1$). Each curve is normalized to its own total. The momentum enhancement curves are the sum of the vector projection of individual momentum (See Method), not sums of individual momentum magnitudes. (**B**) Additive contributions to the Dimorphos momentum enhancement in joint radius and local surface-relative launch-speed bins. Contributions smaller than 0.001 are shown as such. (**C**) Particle fate as functions of radius at T+139.28 days. Stacked areas partition the particle population into material observed in the tail (along the tail with a width of 40 pixels), material escaping the Didymos system without entering the observed tail, material remaining trapped by the binary system, material captured by Dimorphos and Didymos. Horizontal lines mark the global cumulative fractions of the particle number with surface-relative launch speeds less than 0.09, 0.25, 0.5 and 1 m/s and right-axis labels identify the corresponding speed thresholds. Solar radiation pressure and binary gravity filter the ejecta and produce a particle fate that varies non-monotonically with particle size. Within the Didymos system, the fate of ejecta with particle radii below $3 \times 10^{-5}$ m is dominated by radiation pressure, whereas the fate of ejecta with radii between $3 \times 10^{-5}$ and $2 \times 10^{-3}$ m is governed by the coupled effects of radiation pressure and gravity. The particle fate in these two size ranges exhibits a strong

size dependence. For particle radii above $2 \times 10^{-3}$ m, gravity gradually begins to dominate the fate of the ejecta, causing the size dependence of particle fate in this size range to weaken. The pronounced bin-to-bin fluctuations in fate fractions at radii above 1 m reflect the small number of particles in these bins. With such sparse counts, a change in the fate of even one or a few particles can produce a large change in the number-weighted fraction.

### Binary filtering partitions the momentum after launch

After launch, the Didymos binary filtered the ejecta by size and velocity under the combined action of binary gravity and solar radiation pressure (*20*–*22*). The population escaping the Didymos system is therefore a selective subset of the initial ejecta.

Radiation pressure can promote either recapture or escape, depending on ejecta size and where it redirects a trajectory. For sufficiently small particles ($< 3 \times 10^{-5}$ m) launched sunward, strong anti-sunward acceleration rapidly reverses their trajectories, causing them to fall back onto the surface. As particle radius increases, radiation-pressure acceleration decreases ($a_{\mathrm{srp}} \propto r^{-1}$). Consequently, for a given sunward velocity component, both the reversal time and the sunward excursion increase. This increased excursion allows more particles to leave the immediate near-surface environment before being redirected back to the surface by radiation pressure. This weakens the near-surface return pathway of Dimorphos and reduces the fraction of ejecta returning to the surface (Fig. 2C). With further increases in particle radius ($> 3 \times 10^{-5}$ m), the continued weakening of radiation pressure begins to have the opposite effect on capture. For particles that have left the immediate near-surface environment but remain within the binary system, those with larger radii generally have longer residence times, increasing the probability of gravitational deflection and subsequent collision with either body. Consequently, for a given launch speed, the fraction of larger-radius particles that impact Dimorphos or Didymos before radiation pressure drives them out of the system increases with particle radius. This filtering process significantly alters the size–frequency distribution in the dust tail of ejecta with radii between $3 \times 10^{-5}$ m and $2 \times 10^{-3}$ m (Fig. 2C). Consequently, the global size–frequency distribution of the ejecta cannot be described by the three-segment power-law distribution used in previous studies (*9*, *19*) (see Ejecta particle size-frequency distribution in Method, Fig. S2 and table S3 for details).

The filtering process together with dynamical evolution reshapes the momentum budget established at launch by selecting the escaping subset and exchanging momentum, thereby distinguishing momentum generation from momentum transport. The launch field establishes the initial recoil budget, whereas subsequent filtering and dynamical evolution determine which portion of the ejecta is transported outward and in what direction (Figs. 2C and 3). In the representative solution, the binary gravity and solar radiation pressure significantly alter the momentum carried at launch by the ejecta that eventually exit the binary system. This causes the net momentum magnitude of this same particle population to decrease by 23% upon system exit, with its direction rotating by 9° (Fig. 3C). Meanwhile, the momentum projection along the heliocentric orbital-velocity direction decreases by $6.0 \times 10^5$ N s, the corresponding system enhancement decreases from $\Delta\beta_{\mathrm{sys,launch}} = 1.61$ at launch to $\Delta\beta_{\mathrm{sys}} = 1.21$ (Fig. 3B). Within this total change, solar radiation pressure contributes only $-1.5 \times 10^4$ N s. Gravity therefore dominates the direct along-track momentum decrease, while radiation pressure influences momentum direction and the pathways through which particles escape or return.

Ejecta filtering and momentum transport further determine the relationship between $\beta_{\mathrm{D}}$ and $\beta_{\mathrm{sys}}$. In the representative solution, the ejecta escaping the system yield $\beta_{\mathrm{sys}} = 2.21$, and the resulting enhancement ratio κ= $(\beta_{sys} - 1)/(\beta_D - 1)$ = 0.93 summarizes this connection. For all converged solutions, $\beta_{sys} = 2.04 \pm 0.19$ (1σ), κ = $0.92 \pm 0.03$ (1σ) (Fig. 5A). This result lies within the uncertainty of the published astrometric estimate of 2.0 ± 0.3 (*3*). This comparison provides an external consistency check on the ejecta-derived momentum budget with an orbit-side measurement of the same impact.

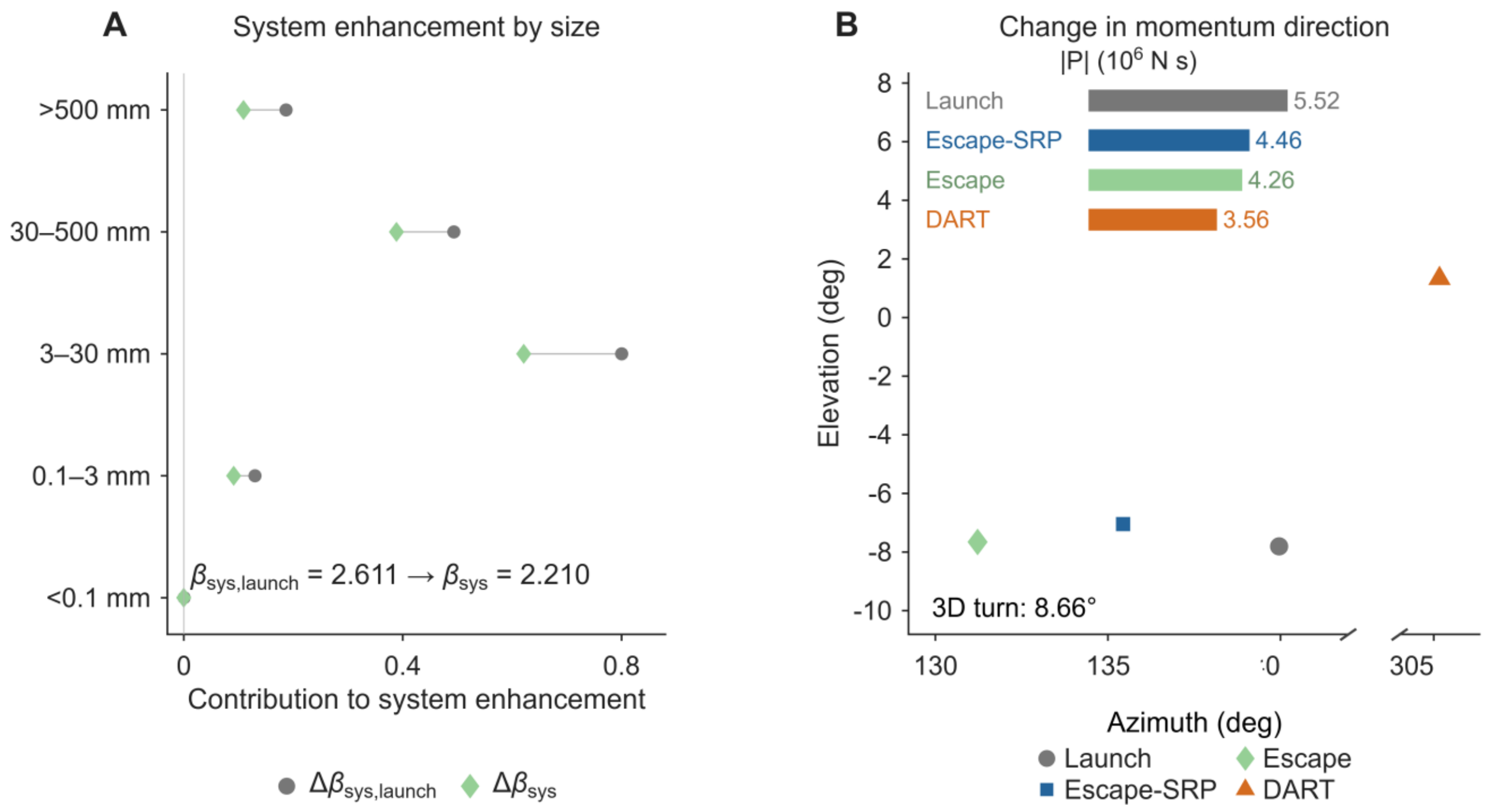

**Fig. 3. Fate selection and momentum transport of the reconstructed ejecta in the representative solution.** (**A**) Size-resolved heliocentric momentum enhancement of the identical system-escape population at launch ($\Delta\beta_{sys,launch} = \beta_{sys,launch} - 1$) (grey circles) and exit ($\Delta\beta_{sys}$) (green diamonds), using each stage's net-direction normalization. (**B**) Change in momentum direction. Azimuth and elevation of the net ejecta momentum for the population escaping the Didymos system at launch (grey circles), at exit (green diamonds), and after subtracting the accumulated solar-radiation-pressure impulse from the escape momentum (blue squares; denoted as Escape-SRP). The orange triangle represents the incident DART momentum. All velocities are measured relative to the binary barycenter, and all vectors are expressed on the nonrotating, right-handed SPICE ECLIPJ2000 axes (*4*). Escape velocities are taken at that particle's recorded outward crossing of the system boundary. Azimuth is defined as $\varphi = \tan^{-1}(P_y/P_x)$, measured from +x towards +y. Elevation is defined as $\theta = \tan^{-1}(P_z/\sqrt{P_x^2 + P_y^2})$, measured from the ecliptic plane towards +z, where $P_x$, $P_y$ and $P_z$ are the vector components along the three coordinate axes. Slashes mark the omitted azimuth interval, and both segments share the same elevation scale. The inset compares the momentum magnitudes, including DART, in units of $10^6$ N s. The annotation "3D turn: 8.66°" indicates the three-dimensional deflection angle of the net momentum vector of the ejecta from launch to escape.

## Bifurcated tail and brightness plateau record the momentum-transport pathways

The bifurcated tail (*9*, *11*, *12*) and brightness plateau (*9*, *12*) revealed in HST observations are diagnostic features of the sorting and filtering process. The bifurcated tail records transport of slow primary ejecta (figs. S5 and S6). Particles with velocities slightly lower than the system escape speed (0.25 m $s^{-1}$) were bent strongly by Didymos gravity, passed over the nightside of Didymos, and were later injected into the tail from the south by solar radiation pressure (fig. S5). Particles with velocities of about 0.25 to 0.3 m $s^{-1}$ passed north of Didymos and entered the tail from the north (fig. S5). These two streams retained their north-south velocity components and expanded into the bifurcated tail observed by HST (fig. S6).

The brightness plateau records a different branch of the same dynamical filtering. Boulders that did not immediately escape the system continued to re-impact the surfaces of Didymos and Dimorphos and produced secondary dust (Fig. 4A), most of which remained temporarily trapped because its velocity was below the system escape speed (*23*). In the representative solution, escaping secondary dust supplied only about 2% of the geometric cross-section of the second tail (Fig. 4A) at T+8.8 days, so it did not build the bifurcated tail (fig. S7). The trapped secondary dust increased the near-system brightness and produced the pause in fading within the 50-km aperture from T + 5.7 to T + 11.9 days (Figs. 4A and B).

Together, these two features show how morphology records momentum partitioning. The bifurcated tail identifies low-velocity primary ejecta routed by binary gravity and solar radiation pressure. The brightness plateau identifies boulder re-impact and secondary dust trapping. Both features trace particles that are essential to the distribution of momentum after impact.

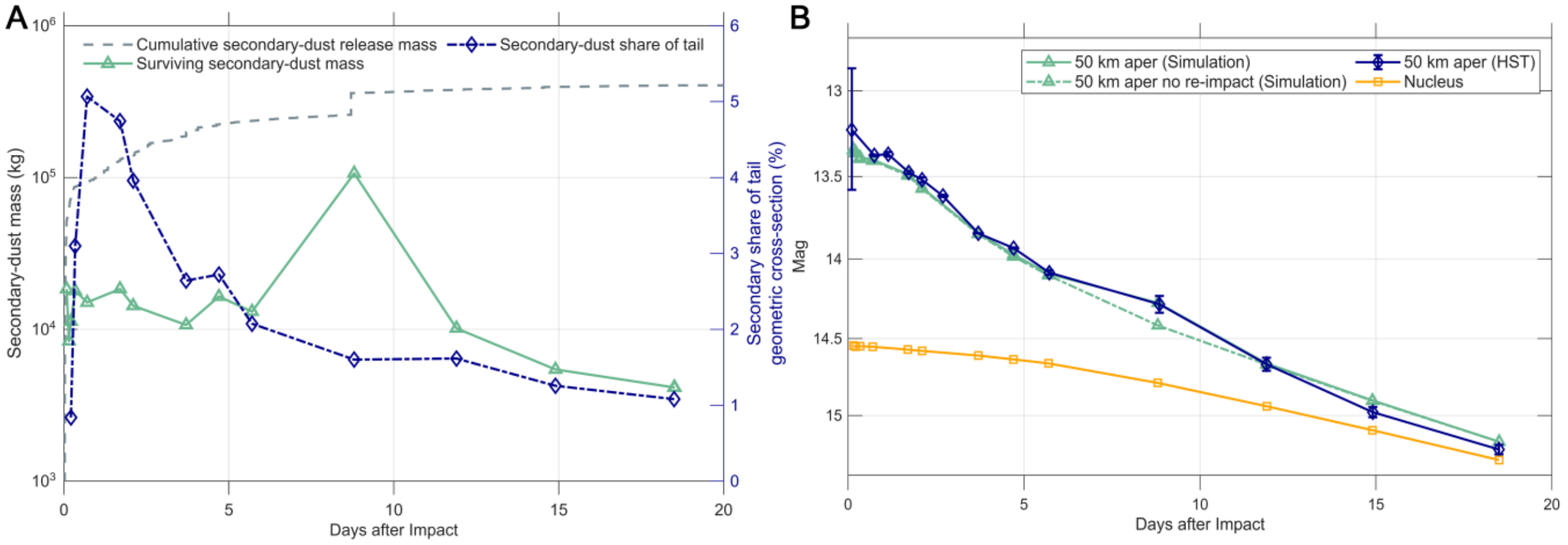

**Fig. 4. Re-impact dust release and nucleus brightness evolution in the representative solution.** (**A**) The grey dashed and green solid curves show the cumulative secondary-dust mass released by ejecta re-impacts and the surviving secondary-dust mass at the indicated epochs, respectively (left logarithmic axis). Surviving secondary-dust mass include both dust that has escaped and dust still trapped in the binary-system. Continued secondary-dust production is offset by reaccretion and system escape, leaving a smaller, temporarily retained near-nucleus dust population. The release event on Didymos at T + 8.7 days increases the near-nucleus surviving dust mass sampled by HST at T + 8.8 days. The dark-blue dash-dotted curve shows the contribution of secondary-dust to the geometric cross-section of the tail at the corresponding observation epoch (right axis). At T + 11.9 d, secondary-dust contribute approximately 2% of the tail cross-section. This geometric cross-section contribution fraction serves as a brightness proxy under common scattering properties. (**B**) Photometry within the 50-km aperture from HST observations (dark-blue diamonds with error bars), simulations including re-impact dust (solid green triangles), and simulations without re-impact dust (dash-dotted green triangles). Orange squares show the nucleus contribution. Retained secondary dust increases the near-system brightness and reduces the fading rate between T + 5.7 and T + 11.9 days, reproducing the observed brightness plateau more closely than the simulation without re-impact dust. The comparison connects the plateau to the re-impact and retention pathway, whereas the bifurcated tail is attributed to escaping primary ejecta. HST aperture photometry is from Extended Data Fig. 4A of ref. (*2*).

## Ejecta recoil constraints the bulk density of Dimorphos

The same momentum budget constrains physical quantities that Hera mission (*24*, *25*) can test. At the nominal along-track velocity change of $\Delta v_T$ =−2.70 mm s$^{-1}$ (*2*), all accepted converged solutions yield $\beta_D$ values of 1.70 to 2.48, $\beta_{sys}$ values of 1.63 to 2.42 and Dimorphos bulk densities of 1139.10 to 1689.98 kg m$^{-3}$ , with respective means of 2.13, 2.04 and 1435.50 kg m$^{-3}$ (Fig. 5A). Propagating the published 1σ uncertainty of 0.10 mm s$^{-1}$ in $\Delta v_T$ (*2*), while holding the remaining ejecta parameters fixed and with equal weighting of the accepted solutions, gives a conditional central 68.3% interval of 1297.84–1597.73 kg m$^{-3}$ for the momentum-inferred density. The inferred transient crater radius across all accepted converged solutions is 32.23 to 48.18 m, and the surviving ejecta mass at T + 18.5 days is $0.73 \times 10^7$ to $1.44 \times 10^7$ kg (Fig. 5B).

These quantities are consequences of the reconstructed momentum budget. Density follows from combining the image-derived recoil with the observed orbital deflection (*2*) and Dimorphos volume (*1*). Crater size and surviving ejecta mass follow from the launch field that reproduces the observed cone, streamers, tail brightness, and boulder population. Hera measurements of Dimorphos mass, crater morphology, surface modification, and boulder distribution will test this source-side reconstruction.

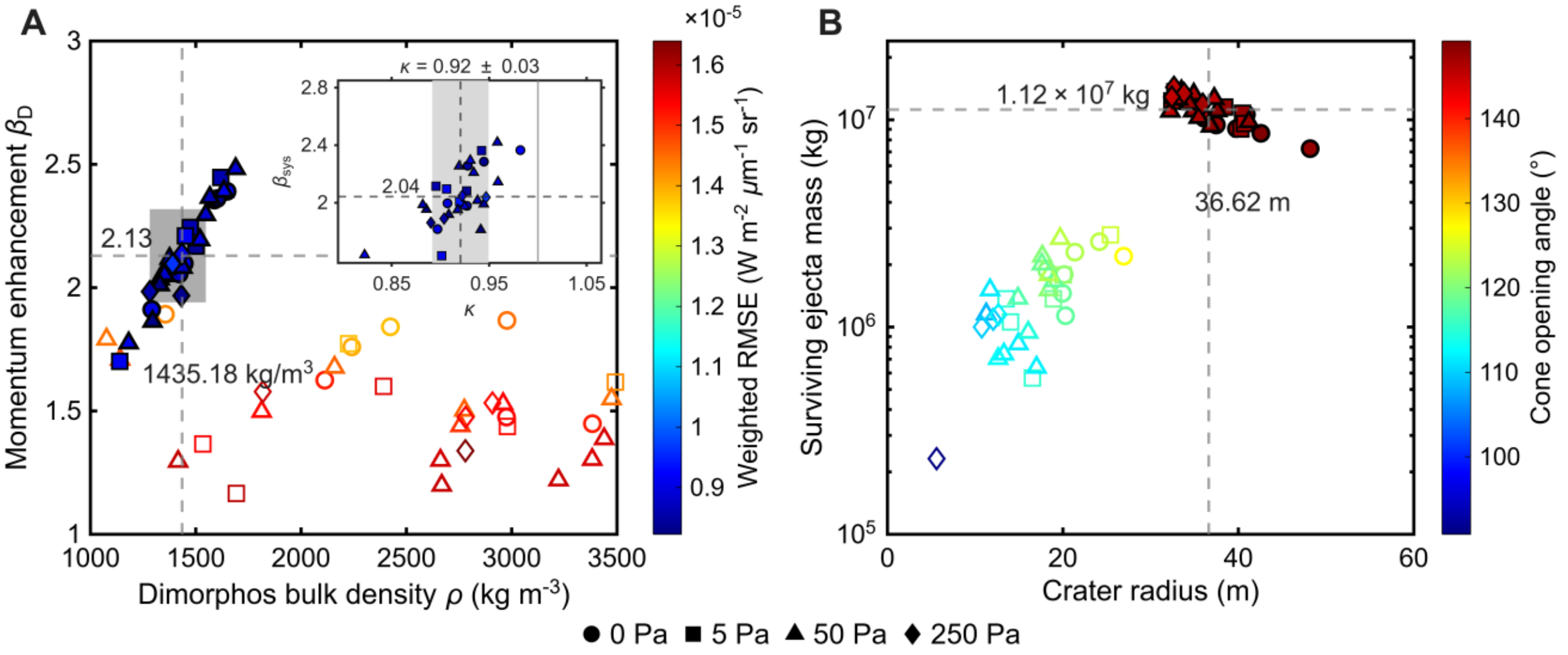

**Fig. 5. Momentum enhancement and physical properties inferred from the ejecta solutions.** (**A**) Dimorphos momentum enhancement, $\beta_D$, versus bulk density. Colors indicate the weighted root mean squared error of the brightness fit. The dashed vertical and horizontal lines, with their corresponding numerical labels, mark the mean converged values of Dimorphos's bulk density and the momentum enhancement factor. The grey region spans the ensemble mean $\beta_D$ ± one sample standard deviation over the densities of accepted solutions within that interval. The upper-right inset shows $\beta_{sys}$ versus $\kappa = (\beta_{sys} - 1)/(\beta_D - 1)$ for the accepted solutions, using the same color scale as the main panel. The dashed vertical line and shaded band indicate the mean κ and ± one sample standard deviation (0.92 ± 0.03). The solid line marks κ = 1. (**B**) Surviving ejecta mass at T + 18.5 days versus inferred transient crater radius. Colors indicate the cone opening angle. Dashed lines in the main panels mark the means of the accepted solutions. Different shapes indicate different cohesive strengths for Dimorphos, with filled and open symbols representing the initial and converged values of the iterative cases, respectively. All quantities are evaluated at the nominal along-track velocity change −2.70 mm s$^{-1}$ (*22*).

## Implications for future impacts and active asteroids

Recovering information about momentum transfer and material transport from the time-resolved morphology and brightness evolution of ejecta links the measured orbital deflection to the material ejection responsible for it, providing a previously missing physical constraint on the ejecta response to impact. This is important for future kinetic-impact missions, as differences in target properties may lead to different ejecta responses and momentum enhancement that cannot be uniquely characterized by orbital measurements alone. The reconstruction also suggests that sufficiently resolved ejecta imaging could constrain recoil before long-term orbital changes are fully characterized.

Moreover, combining the reconstructed recoil with orbital measurements constrains the mass and bulk density of Dimorphos without prescribing its mass beforehand. This extends the information obtainable from impact ejecta beyond the collision itself, providing a way to investigate the physical properties of small bodies whose masses are difficult to measure directly. The inferred density can be independently tested by Hera, while the observation-constrained ejecta field provides a starting point for investigating how continued material transport and re-impacts may have influenced the binary system during the poorly observed interval between the DART and Hera missions.

Beyond planetary defense, the reconstruction provides a controlled reference for investigating the collisional evolution of small Solar System bodies. Natural impacts release material spanning a wide range of particle sizes and velocities, influencing how much material escapes, remains bound, or re-impacts. These different outcomes contribute to collisional mass loss, debris production, and the supply of material to heliocentric orbits, yet the initial ejection conditions are rarely observed directly (*26*). Applying observation-constrained dynamical reconstructions to natural active asteroids could help recover the mass and velocity distributions of released material from its subsequent evolution, while distinguishing separate release episodes from structures produced by later particle transport. Combined with long-term dynamical and collisional models, such constraints could help assess how repeated impacts contribute to material redistribution among small Solar System bodies.

## ACKNOWLEDGMENTS

Y.Y. acknowledges the financial support provided by the National Natural Science Foundation of China Grants No.12272018. F.F. acknowledges support from the European Union ERC Grant agreement No. 101077758 (TRACES) and from the Italian Space Agency (ASI) within the HERA project (agreement n. 2022-8-HH.1-2025). **Funding:** National Natural Science Foundation of China Grants No.12272018 (Y.Y). European Union ERC Grant agreement No. 101077758 (TRACES, F.F). **Author contributions:** Conceptualization: Z.-J.S, Y.Y, J.-Y.L; Methodology: Z.-J.S, Y.Y, J.-Y.L, F.F; Investigation: Z.-J.S, Y.Y, J.-Y.L; Data analysis: Z.-J.S, Y.Y, J.-Y.L; Writing – original draft: Z.-J.S, Y.Y, J.-Y.L, F.F, G.T, P.M; Writing – review & editing: Z.-J.S, Y.Y, J.-Y.L, F.F, G.T, P.M, A.C.B, L.K, T.L.F, A.S.R, B.C, H.B. **Competing interests:** Authors declare that they have no competing interests. **Data, code, and materials availability:** The code and data used to perform the simulations and analysis reported in this work is available from the corresponding author upon request.

## Materials and Methods

Model overview

Figure S1 shows a schematic overview of the numerical model used in this study. The modules shown in the figure are described in detail in the following sections. In brief, the model combines the generation of initial ejecta from an impact on an ellipsoidal surface, ejecta trajectory propagation, discrete-element-method-based simulations of boulder re-impacts, and synthetic image generation (*20*, *27*). Multi-view, multi-epoch observations from LICIACube and HST are used as constraints in a closed-loop iterative optimization of the initial parameter set. The process converges on an optimal set of parameters that yields the corresponding three-dimensional evolutionary history of the ejecta, matching the observations with high fidelity across both spatial and temporal scales.

Compared with existing approaches, this method introduces several advances:

(1) For the first time, our model establishes a closed-loop framework linking the initial parameters governing the three-dimensional development of ejecta to the resulting synthetic images. This connection allows automatic, iterative refinement of the initial parameters based on observational data to obtain the best-fit solution.

(2) Our model captures the complete chain of events in the dynamical evolution of particles, spanning initial ejection and ballistic transport, boulder re-impact and secondary ejecta release, and the subsequent transport of both primary and secondary ejecta;

(3) By simultaneously incorporating multi-source observational data from both HST and LICIACube, our model ensures continuous constraint on the three-dimensional evolution of the ejecta across spatial and temporal scales.

As illustrated in Fig. S1, the model uses the physical properties of Dimorphos (*1*), Maxwell Z-model (*28*, *29*) parameters, and the impact scaling law(*30*) parameters as inputs to generate the initial ejecta mass versus velocity distribution on the surface of Dimorphos. These parameters are categorized as either iteratively updated or fixed, with a complete listing provided in Table S1. The iteratively updated parameters are automatically refined by the model during the simulation until the termination criteria or observational constraints are met. Fixed parameters, in contrast, remain constant throughout a closed-loop iterative process. Together, these two sets of parameters define the full parameter space explored by the numerical simulation (Table S1).

Once the initial ejecta mass versus velocity distribution is generated, the orbital propagator integrates the trajectories of the ejecta particles to obtain their spatiotemporal distribution. During this integration, the model continuously detects whether any ejecta particles re-impact the surface of either component of the binary system. Upon detection of a re-impact event, the model switches from the orbital propagator to a local discrete-element-method (*31*) (DEM) model to simulate the boulder re-impact process. The velocities of the secondary ejecta generated by the re-impact, together with the post-impact velocities of the boulders, are then returned to the orbital propagator to continue the trajectory integration (see DEM-based secondary impacts for details). Finally, the computed spatiotemporal distribution of the ejecta is fed into the synthetic image generation module. Those modules produce simulated outputs (the Dimorphos momentum enhancement factors $\beta_D$ and the heliocentric momentum enhancement factors $\beta_{sys}$, pre-impact Dimorphos's bulk density, synthetic images and photometric data etc.) that can be directly compared with observational data. The discrepancies between these simulated outputs and the actual observations were used as metrics to determine the direction for parameter updates (See Iterative parameter refinement under physical constraints for details), generating improved parameter values for the next iteration. This process continued until all convergence criteria were met or the maximum number of iterations was reached. Notably, the pre-impact bulk density of Dimorphos is a key input parameter required for both the generation of the initial ejecta mass–velocity distribution and the orbital integration. Therefore, this bulk density is treated as an iterative variable: the value inferred from the previous iteration is used to update the input for the subsequent computation, and this process repeats until the residual in bulk density converges (Fig. S1). Using a 32-core Hygon C86 7285 processor (2.0 GHz), a single iteration of the proposed closed-loop model with approximately 1 million particles required ~2 hours, demonstrating that the proposed closed-loop framework is computationally feasible.

Ejecta particle size-frequency distribution

The light scattering properties and the size frequency distribution (SFD) of ejecta strongly influence the reliability of inverting the brightness of ejecta to its mass (*32*). We used the ejecta scattering properties (*33*) derived from theoretical models, laboratory measurements, and observational data to calculate ejecta radiometric flux. In the absence of realistic scattering properties, this allows us to account for dependencies on particle size and phase angle, thereby improving the accuracy of mass estimation.

For the ejecta particle size distribution, pre-impact images of Dimorphos obtained by DART show that boulders with radii larger than 0.5 m on the surface of Dimorphos are well described by a Weibull distribution (*34*). For particles with radii smaller than 0.5 m, however, direct observational constraints are lacking (*34*), leaving the broken power-law distribution inferred from the ejecta tail brightness profiles as the only available reference (*9*, *19*, *27*). The ejecta population may differ from the pre-impact surface population because a substantial fraction of the ejecta may originate from impact-induced fragmentation of boulders (*35*). Furthermore, impact cratering can excavate and eject subsurface material with a size–frequency distribution potentially different from that of surface boulders (*36*). Consequently, the pre-impact surface size distribution may not accurately represent the ejecta particle size distribution.

We model the ejecta particle size distribution using a Weibull distribution for larger particles and a broken power-law distribution for smaller particles, with the distribution parameters subsequently constrained by HST observations. Specifically, we first adopted the Weibull distribution of boulders derived from pre-impact surface images obtained by DART (*34*) and the broken power-law distribution derived from tail-brightness profiles at multiple observational epochs (Table S1) (*9*, *19*) as initial values for iteration. We then iteratively adjust these parameters and the particle-size boundaries between segments until the simulated ejecta brightness distribution matched the observations.

Figure S2 compares the tail brightness profiles extracted from simulated images of the representative solution (Simulation set ID 18 in table S2) with those observed by HST (*9*). The results show that the simulated brightness profile reproduces the observations most closely when the ejecta size–frequency distribution follows a four-segment power law with differential indices of $q_1$ = -3.20 ($1\times10^{-6}$– $3\times10^{-5}$ m), $q_2$ = -2.30 ($3\times10^{-5}$ – $2\times10^{-3}$ m), $q_3$ = -3.18 ($2\times10^{-3}$ mm – $1.3\times10^{-2}$ m), $q_4$ = -4.5 ($1.3\times10^{-2}$ – 1.1 m), together with a Weibull distribution(*34*) with $k_w$= 0.221 and $\lambda_w = 0.0222$ for ejecta larger than 1.1 m (Table S3).

In contrast to the previous results that the ejecta particles from micrometer to millimeter scales follow a single power law (*9*, *19*, *27*), our derived distribution reveals an additional plateau in the SFD in $3\times10^{-5}$ – $2\times10^{-3}$ m range. This feature is required due to a size-filtering process caused by the coupling between solar radiation pressure and the gravitational field of the Didymos binary system, which modifies the size distribution of the ejecta that enters the tail from the global differential size–frequency distribution (dSFD) (Fig. 2C). This size-filtering process, however, only significantly affects the SFD of dust particles in the $3\times10^{-5}$ – $2\times10^{-3}$ m size range (Fig. 2C). Specifically, for ejecta with the intermediate size range of $3\times10^{-5}$ – $2\times10^{-3}$ m, the gravitational force of the Didymos binary system becomes comparable to the solar radiation pressure (*21*, *37*, *38*). The combined effects of gravity and radiation pressure cause a substantial fraction of low-velocity particles in this range to remain gravitationally bound in the binary system or re-impact Dimorphos or Didymos, thereby reducing their abundance in the tail relative to their global proportion, steepening the slope of the SFD. For ejecta larger than $2\times10^{-3}$ m, gravitational force dominates the evolution (*37*), and gravitational scattering injects particles into the tail in a mostly unbiased manner, restoring agreement between the SFD of the tail and the global distributions. With this mechanism in action, reconciling the simulations with observations requires adjusting the global dSFD power-law exponent from −2.7 as reported in the previous study to −2.3 within the $2\times10^{-3}$ m range, in order to compensate for the deficit of grains in the tail caused by gravitational trapping.

Ejecta Initialization

Scaling law-based ejecta plume models have been successfully applied to replicate the evolution of ejecta plumes generated by the NASA Deep Impact (*39*, *40*), Small Carry-on Impactor (SCI) (*41*, *42*) experiment by the JAXA Hayabusa2 mission, and NASA DART impact events (*16*, *43*). However, the determination of the ejection angle of ejecta particles remains inaccessible with scaling laws alone, necessitating additional assumptions regarding their angular distribution (*39*). Experiments (*44*) and numerical simulations (*18*) demonstrate that the Maxwell Z-model, as improved by ref. (*28*) provides a robust description of the evolving morphology of ejecta curtains released form curved surface. Consequently, we establish the ejecta mass versus velocity distribution for DART impact by combining the impact scaling laws with the Maxwell Z-model.

Following steps were took to construct the ejecta mass versus velocity distribution produced by the DART impact:

1) Determination of ejecta mass distribution. Based on scaling laws, the cumulative ejecta mass within a radial distance $x$ from the impact site satisfies the relation (*30*):

$$M(<x) = k\rho_{\text{bulk}}[x^3 - (n_1 a)^3],\ n_1 a \le x \le n_2 R\,, \tag{S1}$$

where $a$ denotes the projectile radius, $\rho_{\text{bulk}}$ represents the bulk density of the target material, and $R$ corresponds to the transient crater radius; $n_1$ and $n_2$ are dimensionless constants. In the gravity-dominated regime, the transient crater radius is expressed as:

$$R\left(\frac{\rho_{\text{bulk}}}{m}\right)^{\frac{1}{3}} = H_1\left(\frac{\rho_{\text{bulk}}}{\delta}\right)^{\frac{2+\mu-6\nu}{3(2+\mu)}}\left(\frac{ga}{U^2}\right)^{-\frac{\mu}{2+\mu}} \tag{S2}$$

where $\delta$ represents the density of the impactor; $\mu$, $\nu$, $H_1$ are dimensionless constants determined by fitting to ejecta mass versus velocity distributions obtained from laboratory experiments or simulations; $U$ and $m$ are the projectile velocity and mass, respectively.

In strength dominated regime, the transient crater radius is given by:

$$R\left(\frac{\rho_{\text{bulk}}}{m}\right)^{\frac{1}{3}} = H_2\left(\frac{\rho_{\text{bulk}}}{\delta}\right)^{\frac{1-3\nu}{3}}\left(\frac{Y}{\rho_{\text{bulk}}U^2}\right)^{-\frac{\mu}{2}} \tag{S3}$$

where $Y$ and $H_2$ are the target strength and dimensionless constants, respectively.

2) Discretization of continuous ejecta mass. Using the size-frequency distribution of ejecta particles, the cumulative ejecta mass derived from scaling laws is discretized into particle clusters via a multiplicative compensation algorithm (See ref. (*20*) in detail). Each cluster is represented by a single weighted particle. To model the secondary dust release caused by re-impacting from ejecta fragments with radii exceeding 0.1 m, originally ejected by the DART impact, we apply direct inverse transform sampling to generate size-frequency distributions with multiplicative compensation factor of 1 for these larger fragments.

3) Spatial allocation of discretized ejecta particles across the Dimorphos surface. A greedy algorithm is employed to randomly distribute ejecta particles across the surface of Dimorphos, centered on the impact point and constrained by the cumulative ejecta mass (Eq. S1). Furthermore, to simulate the ray structures within the ejecta, attributable to boulder blocking (*45*) or impactor shape (*46*), we generate heterogeneous streaks composed of ejecta particles distributed across the Dimorphos surface (Fig. S8B). The positions and widths of these heterogeneous streaks are constrained by LICIACube observations (*8*).

4) Assigning ejecta velocities and directions. Ejecta kinematics (velocity magnitude and direction) originating from launch position $x$ on Dimorphos surface is determined by scaling laws and the Maxwell Z-model. The velocity magnitude $v_{\mathrm{ej}}$ of an ejecta particle at a radial distance $x$ is governed by the scaling relationship:

$$v_{\mathrm{ej}} = C_1 U \left[\frac{x}{a}\left(\frac{\rho_{\mathrm{bulk}}}{\delta}\right)^{\nu}\right]^{-\frac{1}{\mu}} \left(1-\frac{x}{n_2 R}\right)^{p}, n_1 a \le x \le n_2 R \, , \tag{S4}$$

where $C_1$ and $p$ are scaling constants. The angle $\theta$ (Fig. S1A) between the velocity vector and the line connecting the flow center to the ejection position on the target surface is determined by the improved Maxwell Z-model (*28*):

$$\tan\theta = (\mathrm{Z}-2)\frac{\sin\phi}{1+\cos\phi} \tag{S5}$$

Here, $\phi$ is an angle from the vertical axis (Fig. S8A); Z defines how sharply the streamline changes its direction. Consistent with impact-cratering flow studies (*47*, *48*), we prescribe a Z value that increases linearly with launch position $x$ and place the flow center at depth *d* (Fig. S8A), which best reproduces the ejecta curtain morphology. All parameter values used in each simulation are listed in Table S1.

<u>Orbital dynamics propagator</u>

In our model, the ejecta is assumed to consist of homogeneous, spherical particles that propagate without inter-particle collisions or exerting any dynamical effects on the binary system, because such effects, if any, tend to occur in a much longer timeframe than what we analyze here. Moreover, since tidal perturbations induced by planets are at least 2 orders of magnitude weaker than the solar tidal effect over the simulated time span (*21*), they are neglected. Consequently, once ejected from the asteroid surface, the trajectory of each particle is governed solely by the combined effects of the binary system's gravity, solar radiation pressure (SRP), and the solar tide. The equation of motion for an ejected particle, expressed in a quasi-inertial reference frame centered at the barycenter of the Didymos binary system, with the *x*-, *y*- and *z*-axes aligned with those of the heliocentric ecliptic J2000 coordinate system, can be written as(*20, 21*):

$$\ddot{\boldsymbol{r}}_{\mathrm{p}} = -GM_{\mathrm{s}}\left(\frac{(\boldsymbol{r}_{\mathrm{p}}-\boldsymbol{r}_{\mathrm{d}})}{\left|\boldsymbol{r}_{\mathrm{p}}-\boldsymbol{r}_{\mathrm{d}}\right|^3}+\frac{\boldsymbol{r}_{\mathrm{d}}}{\left|\boldsymbol{r}_{\mathrm{d}}\right|^3}\right) - GM_{\odot}\left(\frac{(1-\beta_{\mathrm{spr}})(\boldsymbol{r}_{\mathrm{p}}-\boldsymbol{r}_{\mathrm{s}})}{\left|\boldsymbol{r}_{\mathrm{p}}-\boldsymbol{r}_{\mathrm{s}}\right|^3}+\frac{\boldsymbol{r}_{\mathrm{s}}}{\left|\boldsymbol{r}_{\mathrm{s}}\right|^3}\right) + \nabla U_{\mathrm{p}} \tag{S6}$$

where $\boldsymbol{r}_{\mathrm{p}}$, $\boldsymbol{r}_{\mathrm{d}}$, and $\boldsymbol{r}_{\mathrm{s}}$ are the position vectors of the particle, Dimorphos, and the Sun, respectively; $M_{\odot}$ is the mass of the Sun; $M_{\mathrm{s}}$ is the mass of Dimorphos, which is determined by iterative convergence value; $\beta_{\mathrm{spr}} = \frac{1.141\times10^{27}\varepsilon}{M_{\odot}\rho r}$, where $\rho = 3500$ kg m$^{-3}$ is the grain density matching the L and LL chondrites (*49*), which are the best meteorite analogs for Didymos (*50*), assuming Dimorphos shares the same composition; *r* is the radius of the ejected particle and $\varepsilon = 1.16$ is the reflection coefficient of the particle (*21, 51*); $\nabla U_{\mathrm{p}}$ indicates the gravity from the primary. For Didymos, we assume it as a homogeneous polyhedron and using the Didymos v003 shape model (*1*) to represent its shape. Then, the gravity $\nabla U_{\mathrm{p}}$ from Didymos can be evaluated by the polyhedral method (*52*). For Dimorphos, its shape is assumed to be a triaxial ellipsoid (*1*) but its gravity is represented by a point mass.

Solar occultations caused by the binary system occur frequently for ejecta particles orbiting nearby, leading to repeated on–off cycles of the solar radiation pressure term in the equations of motion (*21*). Therefore, a fast solar occultation detection algorithm (*21*) is incorporated into the orbital integration. The ephemerides of all relevant celestial bodies are obtained from NASA's DART mission SPICE kernels (*4*) in our model. We numerically propagate the orbital dynamics of the particles using the adaptive Runge-Kutta-Fehlberg (RKF78) method (*53*), with relative and absolute tolerances both set to $10^{-13}$.

DEM-based secondary impacts

Re-impact events between the original ejecta released by the DART impact and the asteroid surface represent interactions between incident particles and regolith particles. The DEM provides a suitable computational approach for modeling such interactions (*54*). However, given the computational cost of DEM simulations, it is infeasible to model the re-impacts of all ejecta fragments. Consequently, DEM simulations were only used to model re-impacts of particles with radii larger than 0.1 m. For smaller ejecta particles (radius < 0.1 m), we model the re-impacts as a collision between a sphere and a flat plane using a classic collision theory of rigid bodies (*55*). This approach excludes any secondary dust contribution resulting from ejecta particles with radius smaller than 0.1 m undergoing re-impacts.

The algorithm from ref. (*54*) enables collision detection between particles and the surface at every orbital integration timestep. It rapidly determines both the collision location and the corresponding surface normal vector $\boldsymbol{N}$ in the global frame. Upon detecting a secondary impact event caused by a large particle, the model switches from orbital integration to a local DEM simulation of the re-impact event. This transition enables simulation of particle-surface collision dynamics at the re-impact site.

The local frame for the DEM substrate was defined with its vertical axis $\boldsymbol{N}_{local}$ normal to the surface plane (Fig. S1). The transformation matrix $\boldsymbol{T}$ between global to local frames was constructed using Rodrigues' rotation formula (*56*):

$$\boldsymbol{T} = \mathbf{I} + \sin\theta \cdot \boldsymbol{K} + (1\text{-}\cos\theta) \cdot \boldsymbol{K}^2, \qquad \text{(S7)}$$

where $\theta$ denotes the angle between the asteroid's surface normal $\boldsymbol{N}$ and the substrate surface normal $\boldsymbol{N}_{\mathrm{local}}$, $\mathbf{I}$ is the identity matrix, and $\boldsymbol{K}$ is the skew-symmetric matrix for the rotation axis $\boldsymbol{k} = \frac{\boldsymbol{N}\times\boldsymbol{N}_{\mathrm{local}}}{\left\|\boldsymbol{N}\times\boldsymbol{N}_{\mathrm{local}}\right\|}$. Using matrix $\boldsymbol{T}$, both the incident velocity vector and gravitational acceleration vector were transformed from the global frame to the DEM substrate's local frame.

Since the computational efficiency of the DEM degrades rapidly with increasing particle–size disparity, we assumed that re-impacts occur on monodispersed granular substrates composed of four distinct grain sizes—0.07 m, 0.21 m, 0.63 m, and 1.9 m—to improve the computational efficiency of our model. Each substrate was pre-prepared under microgravity ($g=10^{-5}$ m s$^{-2}$) via free-fall deposition. Our model automatically selected the substrate with the closest smaller grain size relative to the re-impacting particle radius as the DEM substrate. For example, re-impacts by particles with radii between 0.21 and 0.63 m radius would trigger a selection of the 0.21 m substrate.

Following initialization of the local incident velocity, these prepared substrates underwent a secondary settling process under the local effective gravitational acceleration (the gravitational acceleration plus the centrifugal acceleration). Once settling was complete, the DEM simulation commenced automatically. We adopted a bulk density of 2,780 kg m$^{-3}$ for Didymos to prevent the local effective gravitational acceleration from pointing outward on its surface (*57*). Considering that some secondary impacts locations might lie on steep slopes where substrate settling is not feasible, these sites were treated as exposed boulders, and no ejecta were generated from secondary impacts at these locations.

Secondary ejecta particles were identified and recorded when: (i) their velocity vectors formed acute angles with the surface normal $\boldsymbol{N}_{\text{local}}$; and (ii) their position exceeds one particle radius above the original substrate surface. The DEM simulation terminated when the total kinetic energy of the particle bed (excluding secondary ejecta particles) decayed below $10^{-4}$ times the initial incident particle kinetic energy or the average velocity of substrate particles below 0.001 m s$^{-1}$. All identified secondary ejecta particles were discretized into micrometer to centimeter scale particles following the ejecta size–frequency distribution (Fig. S2), while conserving both total mass and the velocity distribution. These secondary ejecta particles, together with the originally incident particle, updated to its post-impact state, were then transformed into the global frame. The model subsequently resumed orbital propagation for the incident particle and all secondary ejecta (Fig. S1).

Synthetic image and aperture photometry

Synthetic images and photometric data corresponding to the observational conditions of HST and LICIACube were generated following the same technique as in ref (*20*) and ref (*33*), respectively. LICIACube's trajectory was derived from NASA's DART mission SPICE kernels (*4*), while HST ephemeris data were obtained from JPL Horizons. For LICIACube synthetic images, the camera was set to point at Dimorphos with a focal length (*6*) of 70.55 mm, instantaneous field of view(*6*) (IFOV) of 78 μrad. For HST synthetic images, Didymos was modeled as a sphere with radius of 390 m. The camera pointed toward the system's barycenter with focal length (*58*) 57.6 m, IFOV (*58*) of 0.194 μrad. The ejecta scattering properties in ref (*33*) were used to compute the radiometric flux, which allows us to account for dependencies on ejecta size and phase angle. For HST simulated images, the apparent magnitude $m_p$ of individual ejecta particles was calculated as:

$$m_{\text{p}} = m_{\text{sun}} + 5\log_{10}\left(d_{\text{P,C}}d_{\text{P,S}}\right) - 2.5\log_{10}\left(A_{\text{p}}(a, r_{\text{p}})r_{\text{p}}^2\right) \tag{S8}$$

where $d_{\text{P,C}}$ and $d_{\text{P,S}}$ are particle–camera and particle–Sun distances, respectively; $m_{\text{sun}} = -26.76$ is the apparent magnitude of the Sun at 1 AU; $A_{\text{p}}(a, r_{\text{p}})$ is the phase-dependent geometric albedo of a spherical particle with radius $r_{\text{p}}$ and phase angle $\alpha$. Using the Vega reference system, the radiometric flux of a single particle is:

$$E_{\mathrm{p}} = E_{\mathrm{Vega}} 10^{(\frac{m_{\mathrm{Vega}} - m_{\mathrm{p}}}{2.5})} \quad \text{(S9)}$$

where $m_{\mathrm{Vega}} = 0.03$ is the Vega magnitude(*20*), $E_{\mathrm{Vega}} = 3.631 \times 10^{-8}$ W (m$^2$ nm)$^{-1}$ is Vega reference irradiance in the visible (*20*). The radiometric flux of a single particle combined with the multiplicative compensation factors, yielded the total radiometric flux for particle clusters represented by individual weighted particle (*20*). Then per-pixel fluxes were obtained by summing all weighted fluxes projected onto same pixel coordinates. Finally, the per-pixel results were converted to the same units as HST data to produce the simulation image using the following calculation (*20*):

$$I = \frac{E}{IFOV^2} \frac{E_{\mathrm{Filter}}}{E_{\mathrm{Vega}}} \quad \text{(S10)}$$

where $E$ is the rendered image, $I$ is the image with same units as HST data, and $E_{\mathrm{Filter}}$ = 2.7554 × $10^{-8}$ W (m$^2$ nm)$^{-1}$ is the F350LP filter zero point.

For LICIACube simulated images, the specific intensity of a particle with radius $r_p$ was calculated as:

$$I = \frac{A_{\mathrm{p}}(a, r_{\mathrm{p}}) \pi r_{\mathrm{p}}^2 J_{\odot}}{d_{\mathrm{P,C}}^2 IFOV^2} \quad \text{(S11)}$$

where $J_{\odot} = 1.65$ W m$^{-2}$ nm$^{-1}$ is the filter weighted solar flux in the LUKE R channel(*33*).

The optical depth along a given line of sight is computed as:

$$\tau = \int \sigma_{\mathrm{ext}}(\mathrm{s}) \mathrm{ds} \quad \text{(S12)}$$

The volume extinction coefficient along the line of sight $\sigma_{\mathrm{ext}}$ is evaluated on a voxel basis as the sum of weighted particle extinction cross sections divided by the voxel volume:

$$\sigma_{\mathrm{ext}} = \frac{\sum_i Q \pi r_{\mathrm{p},i}^2}{\Delta V} \quad \text{(S13)}$$

where $Q$ = 2 is the extinction efficiency (*16*) and $\Delta V$ is the voxel volume.

Momentum-enhancement factors

In kinetic impact deflection missions, the momentum acquired by the target asteroid, the impactor incident momentum, and the ejecta's net momentum satisfy the momentum balance,

$$M_{\mathrm{post}} \Delta \boldsymbol{v} = m_{\mathrm{dart}} \boldsymbol{U} - \boldsymbol{p}_{\mathrm{ej}} \quad \text{(S14)}$$

where $M_{\mathrm{post}}$ is the Dimorphos mass after impact, $\Delta \boldsymbol{v}$ is the change in Dimorphos's orbital velocity induced by the impact, $m_{\mathrm{dart}}$ is the spacecraft mass at time of impact, $\boldsymbol{U}$ is the impact velocity vector relative to Dimorphos and $\boldsymbol{p}_{\mathrm{ej}}$ is the net momentum of ejecta escaping from Dimorphos. The latter is defined as the vector sum of the momentum of all particles escaping the gravitational influence of Dimorphos:

$$\boldsymbol{p}_{\mathrm{ej}} = \sum_{\text{all esp}} m_{\mathrm{ej}} \boldsymbol{v}_{\mathrm{ej}} \quad \text{(S15)}$$

where $m_{ej}$ represents the mass of an ejecta particle escaping from Dimorphos and $\boldsymbol{v}_{ej}$ its velocity vector. The momentum enhancement factor for Dimorphos $\beta_D$ is defined by the momentum balance of the kinetic impact (*2*, *14*),

$$\beta_D = 1 - \frac{\boldsymbol{p}_{ej} \bullet \mathbf{e}_T}{m_{dart}(\boldsymbol{U} \bullet \boldsymbol{E})(\boldsymbol{E} \bullet \mathbf{e}_T)} \tag{S16}$$

where $\boldsymbol{E}$ and $\mathbf{e}_T$ are the net ejecta momentum direction and the direction of Dimorphos's orbital motion, respectively.

Once the ejecta momentum vector $\boldsymbol{p}_{ej}$ is determined, the Dimorphos mass before impact $M_s$ can be derived from:

$$M_s = \frac{m_{dart}\boldsymbol{U} \bullet \mathbf{e}_T - \boldsymbol{p}_{ej} \bullet \mathbf{e}_T}{\Delta \boldsymbol{v} \bullet \mathbf{e}_T} + \sum_{\text{all esp}} m_{ej} - m_{dart} \tag{S17}$$

where $\Delta \boldsymbol{v} \bullet \mathbf{e}_T = -2.7 \text{ mm s}^{-1}$ represents the along-track component of Dimorphos's orbital velocity change (*2*). The bulk density of Dimorphos $\rho_{bulk}$ is then calculated as:

$$\rho_{bulk} = \frac{M_s}{V} \tag{S18}$$

where $V = 1.81 \times 10^6 \text{ m}^3$ is the Dimorphos volume (*1*).

The heliocentric momentum enhancement factor is defined as:

$$\beta_{sys} = 1 - \frac{\boldsymbol{p}_{sej} \bullet \boldsymbol{e}_{sys,T}}{m_{dart}(\boldsymbol{U}_{sys} \bullet \boldsymbol{E}_{sys})(\boldsymbol{E}_{sys} \bullet \boldsymbol{e}_{sys,T})} \tag{S19}$$

where $\boldsymbol{p}_{sej}$ is the net momentum of ejecta escaping from Didymos system, $\boldsymbol{E}_{sys}$ is the net momentum direction of ejecta escaping from Didymos system, $\boldsymbol{e}_{sys,T}$ is the direction of the system barycenter orbital motion and $\boldsymbol{U}_{sys}$ is the impact velocity vector relative to the system barycenter.

Iterative parameter refinement under physical constraints

We constrain the physical properties of Dimorphos and iteratively refine the Maxwell Z-model and scaling law parameters using image data from HST and LICIACube. The convergence criteria and parameter refinement logic in our simulations are as follows:

1) The absolute difference between the ejecta cone opening angle in the synthetic images and that measured in the LICIACube observations does not exceed 2.5° at T + 174 s (Fig. S1). If the opening angle extracted from the synthetic image is larger than that observed, the parameter $Z$ at $x$=0 in the Maxwell Z-model is decreased; otherwise, $Z$ is increased.

2) The residual in Dimorphos's bulk density must be less than 30 kg $m^{-3}$, corresponding to a residual in Dimorphos's mass approximately 1% of its total mass, which is in a comparable order of magnitude as the total ejecta mass.

3) Tail brightness profiles agreement: For the linear parts of the tail brightness profiles, the absolute differences in slope and distance between the synthetic and observed images do not exceed 0.1 and 10 pixels, respectively.

4) Radiometric flux agreement: The weighted root mean squared error (RMSE) of radiometric radiance between synthetic and observed images (outside a 50-km photometric aperture) does not exceed $1.2\times10^{-5}$ W ($m^2$ μm sr)$^{-1}$. Additionally, the RMSE of the ejecta brightness curve over time (measured within a 50 km photometric aperture) between the simulated and observed results must be below 0.15 magnitudes. The two convergence criteria were chosen to balance computational cost against the need to maximize agreement between simulated and observed images. The weights assigned to the data at each observation time (Table S4) were defined based on the magnitude of the observational uncertainties (Extended Data Figure 4b of ref. (*9*)). Because the ejecta within the 50 km aperture appears to be almost featureless in both simulations and observations, its total brightness provides a sufficient constraint. For ejecta outside the 50 km aperture, brightness alone is insufficient to capture structural discrepancies between simulations and observations. Therefore, a weighted RMSE that simultaneously accounts for both brightness and morphology is employed to quantify the quality of simulation for the outer ejecta.

To achieve convergence of the model, we adopted the following parameter refinement strategy. The brightness of ejecta within the 50 km photometric aperture is used as the primary metric to update the scaling law parameter $\mu$, while the RMSE of the radiometric flux between the synthetic and observed images guides the refinement of the scaling parameters $H_1$ (or $H_2$, depending on the cohesion strength) and $C_1$. Specifically, if the simulated brightness within the 50 km aperture exceeds the observations, $\mu$ is reduced; otherwise, it is increased. For $H_1$ and $C_1$, a direct search method is used to determine the adjustment direction. A small perturbation is applied to the initial value of $C_1$ and evaluate the resulting change in RMSE. If the RMSE decreases, $C_1$ is updated in the perturbation direction; otherwise, it is updated in the opposite direction. If $C_1$ reaches its upper or lower bound (Table S1), the same search strategy is then applied to update $H_1$.

In our simulations, if all adjustable parameters reach their prescribed bounds or if convergence is not achieved within 50 iterations, the case corresponding to the initial conditions is classified as non-convergent and discarded. Furthermore, a converged solution is deemed invalid if it violates any of the following physical constraints:

1) Total Ejecta Kinetic Energy Limit: The total kinetic energy of the ejecta must not exceed the incident kinetic energy delivered by the DART impactor.

2) Positive Macroporosity Requirement: The derived macro-porosity $\alpha_D$ of Dimorphos must be greater than zero, implying an upper limit for the mass of Dimorphos.

3) Momentum Enhancement Consistency: The calculated momentum enhancement factor $\beta$ must be greater than one.

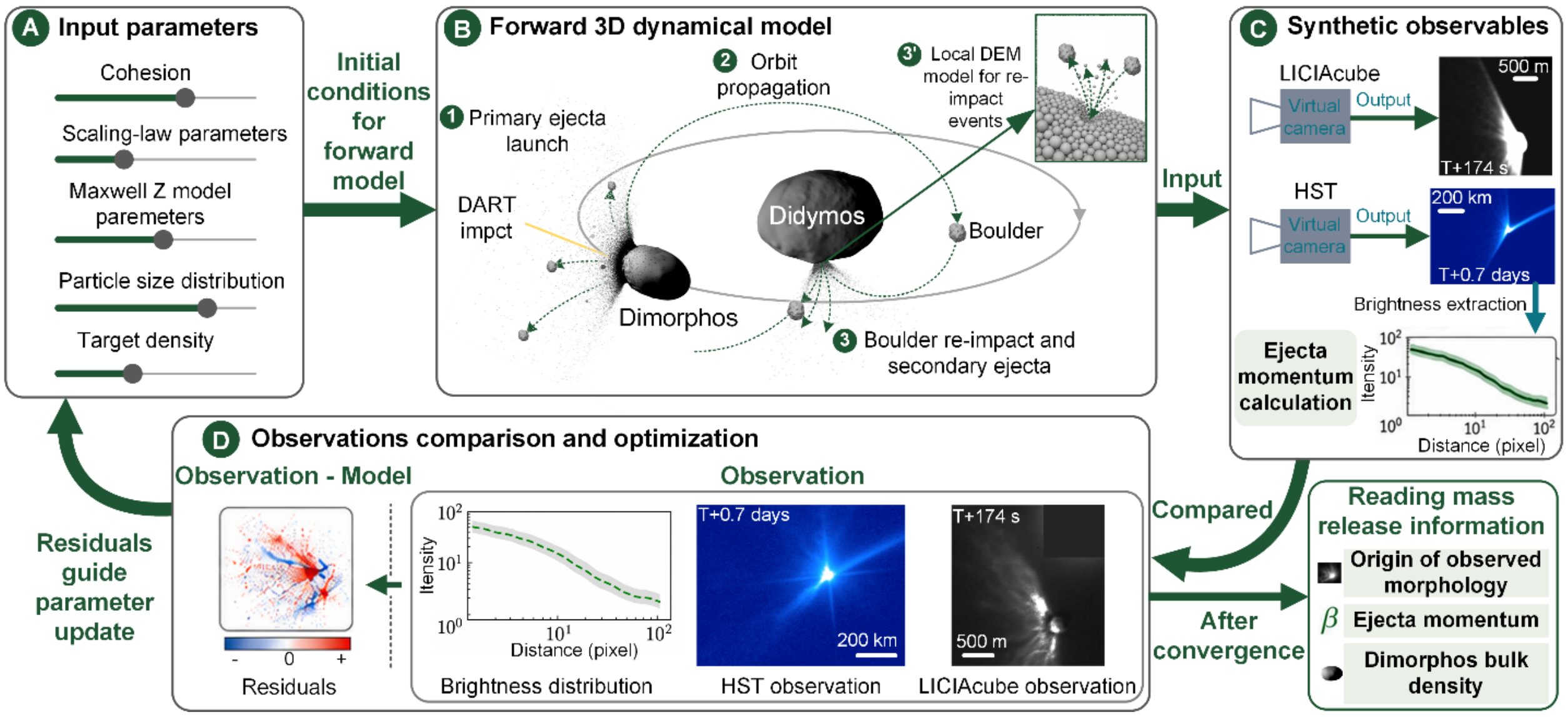

**Fig. S1.**

Schematic overview of the numerical model. The model is initialized with key parameters including material cohesion strength, impact scaling-law parameters, Maxwell Z-model parameters, the ejecta size-frequency distribution and Dimorphos's bulk density. The Ejecta Initialization Module (**A**) uses these inputs to generate the ejection positions and velocities of ejecta particles from the surface of Dimorphos (see Ejecta Initialization for details). Forward 3D dynamical model Module (**B**) integrates the initial conditions generated by Module (**A**) and detects whether primary ejecta impact the surfaces of either binary components (see Orbital dynamics propagator for details). If an impact occurs, it performs local DEM modeling of the impact process for boulders with radii greater than 0.1 m and incorporates the secondary ejecta produced by that impact event and the rebounded boulders into the orbital propagation (see DEM-based secondary impacts for details). Synthetic observables Module (**C**) takes the spatiotemporal distribution of ejecta output by Module (**B**) as input and converts the three-dimensional distribution of ejecta into quantities directly comparable to observations (see Synthetic image and aperture photometry for details). Observations comparison and optimization Module (**D**) directly compares the simulation results output by Module (**C**) with actual observations from HST and LICIACube. The resulting deviations between simulations and observations are used to guide updates to the input parameters of Module (**A**) for the next run (see Iterative parameter refinement under physical constraints for details). Once the iterative process converges, a reconstructed ejecta solution consistent with observations is obtained. This solution can be used for subsequent work such as evaluating kinetic deflection effectiveness and interpreting the momentum partitioning of ejecta.

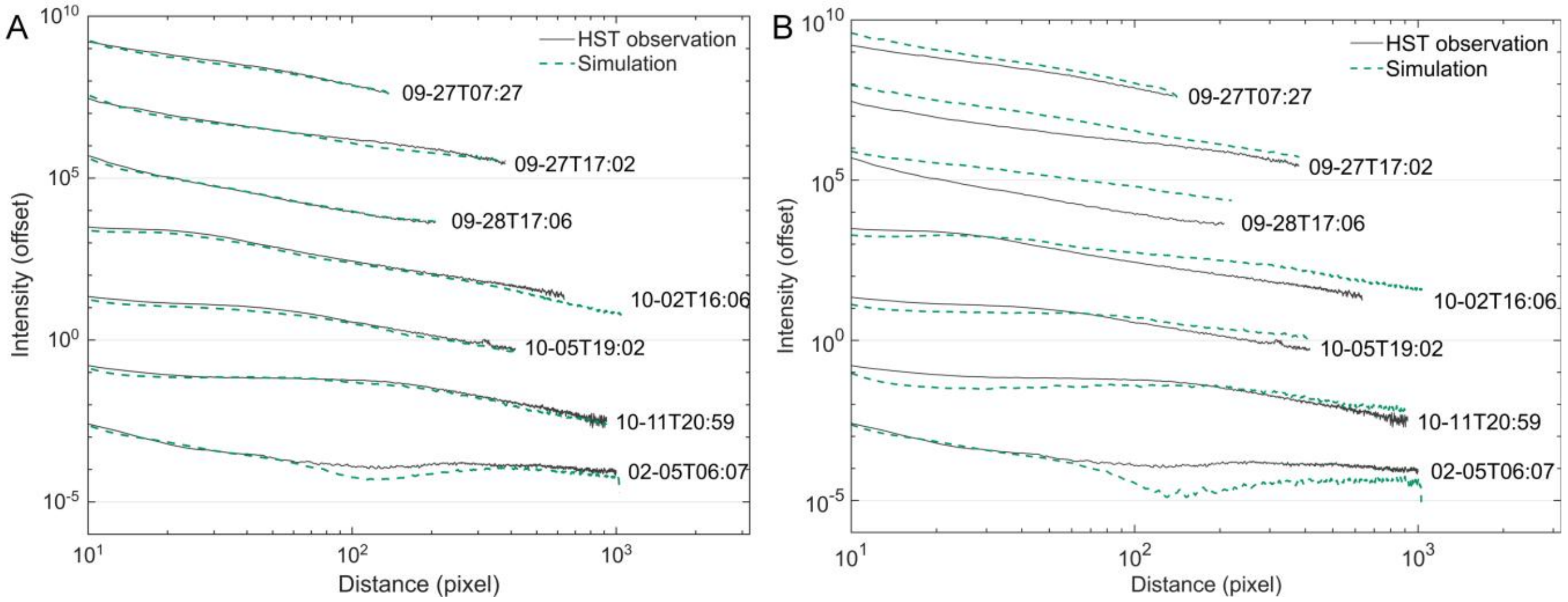

**Fig. S2.**

Tail brightness profiles from HST observations compared with numerical simulations. The green dashed and black solid lines represent the average surface brightness extracted along the tail (width = 40 pixels) from the numerical simulations and HST observations, respectively, and are vertically offset by the same amount for clarity. (**A**) Tail brightness profile corresponding to a four-segment power-law distribution with differential indices $q_1$ = -3.20 ($1\times10^{-6}$– $3\times10^{-5}$ m), $q_2$ = -2.30 ($3\times10^{-5}$ – $2\times10^{-3}$ m), $q_3$ = -3.18 ($2\times10^{-3}$ mm – $1.3\times10^{-2}$ m), $q_4$ = -4.50 ($1.3\times10^{-2}$ – 1.1 m), combined with a Weibull distribution with $k$= 0.221 and $\lambda = 0.0222$ for ejecta larger than 1.1 m(*34*), compared with the tail brightness profile extracted from HST observations(*9*). (**B**) Tail brightness profile corresponding to a three-segment power-law distribution (*19*) with differential indices $q_1$=−2.7 (1 µm–2 mm), $q_2$ =−3.7 (2 mm–1 cm), and $q_3$ =−4.2 (1 cm–0.5 m), combined with a same Weibull distribution for ejecta larger than 0.5 m (*34*). The four-segment power-law distribution reproduces both the observed tail brightness profile, whereas the three-segment power-law distribution is inconsistent with the observations (*9*).

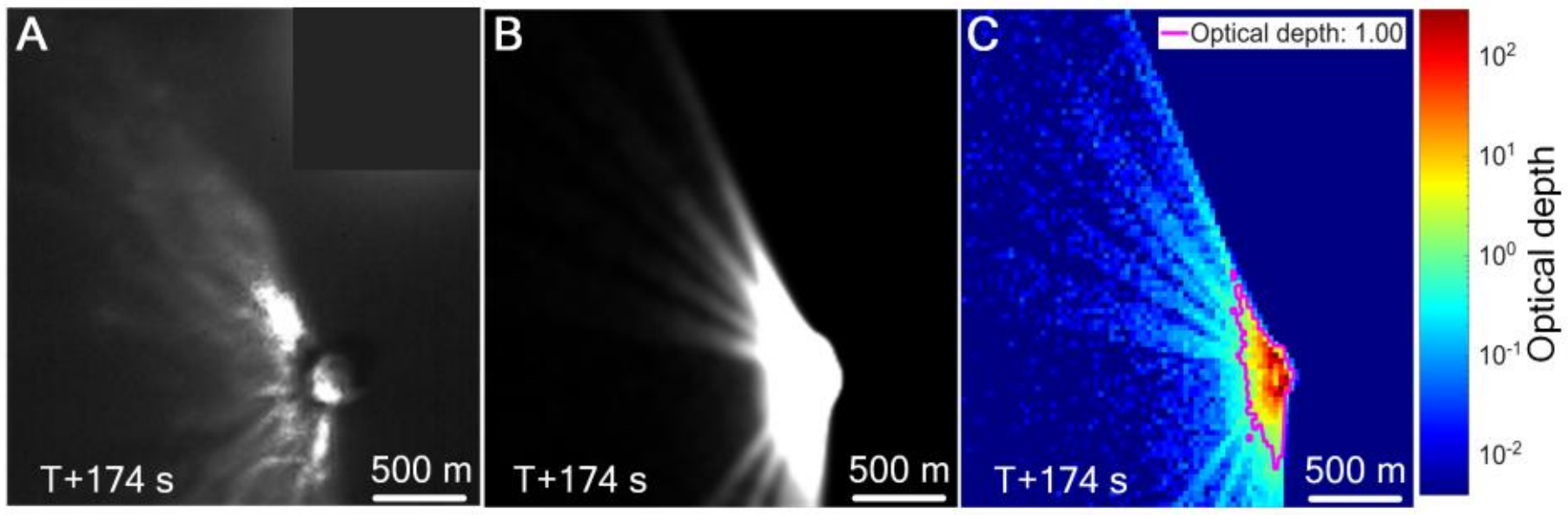

**Fig. S3.**

Optical depth distribution of the ejecta. (**A**) Image of the ejecta cone acquired by the LUKE at a distance of 71 km, 6.5 s after closest approach (174 s after impact). It is adapted from the calibrated red filter image available in the Planetary Data System (PDS). (**B**) Synthetic images from numerical simulation under same acquisition time, viewing geometry, and illumination conditions as the LUKE image. (**C**) Optical depth distribution of the ejecta. The purple solid line denotes the contour corresponding to an optical depth of 1. This optically thick cloud casts a shadow on the material beneath (panel **A**).

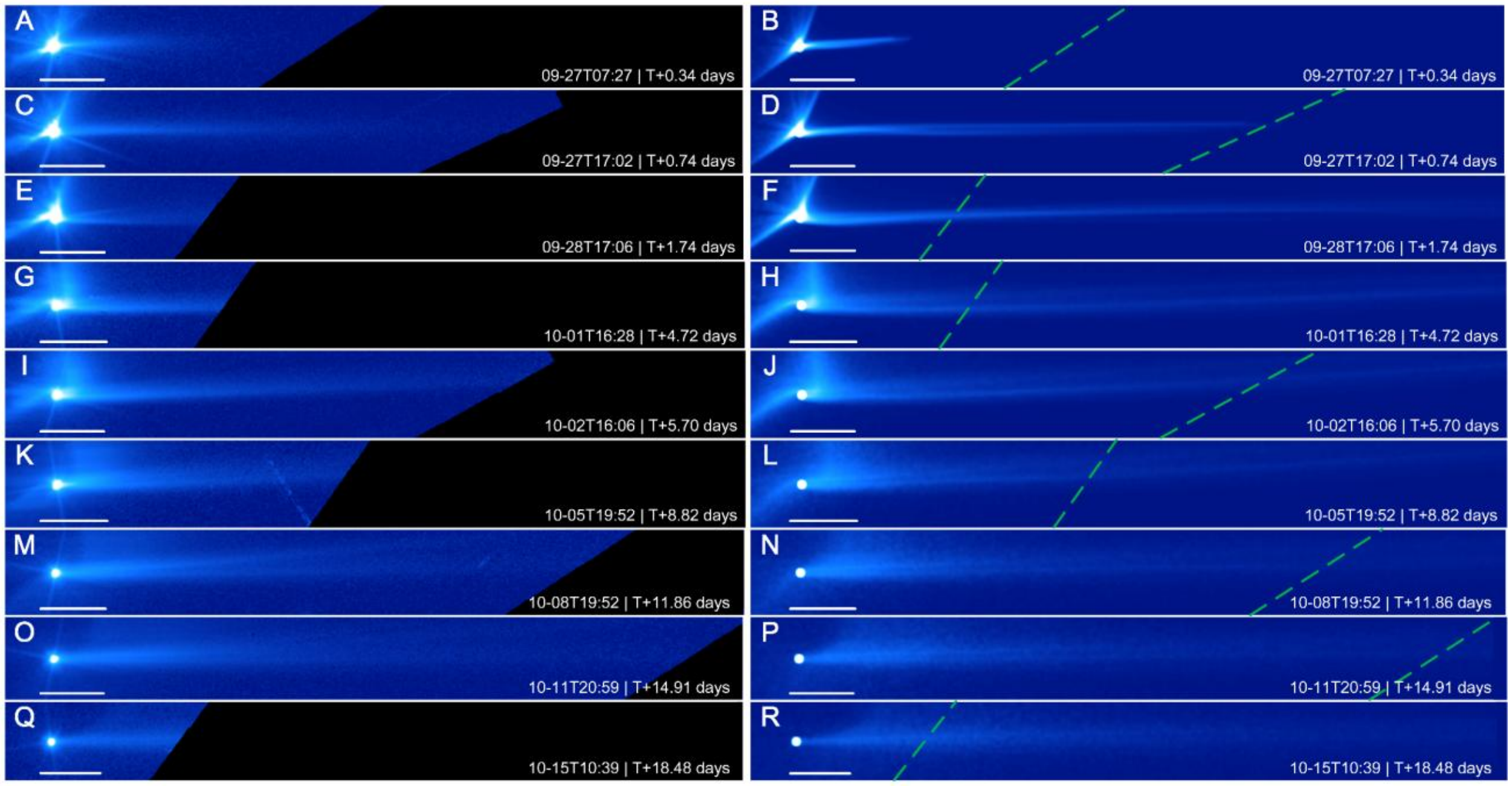

**Fig. S4.**

Comparison of HST images with synthetic images from numerical simulation of ejecta tail formation. All images are displayed with the same logarithmic brightness scale and stretch. Scale bars are 200 km at the distance of Didymos. The left panels (**A**, **C**, **E** , **G**, **I**, **K**, **M**, **O**, **P** ) show HST images (*9*), with corresponding synthetic images (**B**, **D**, **F**, **H**, **J**, **L**, **N**, **P**, **R**) from representative numerical simulations in the right panels. All panels are rotated such that the ejecta tails are oriented nearly horizontally. In the synthetic images, green dashed lines mark the boundaries of the HST field of view. The secondary tail, morphologically thinner than the primary tail, began to appear north of the original tail at approximately T+4.72 days (**H**). It reached a width comparable to that of the primary tail within the HST field of view by T + 8.82 days (**L**) and subsequently disappeared from the FOV by T + 18.48 days (**R**). Panels **A**, **C**, **E**, **G**, **I**, **K**, **M**, **O**, and P are adapted from ref. (*9*).

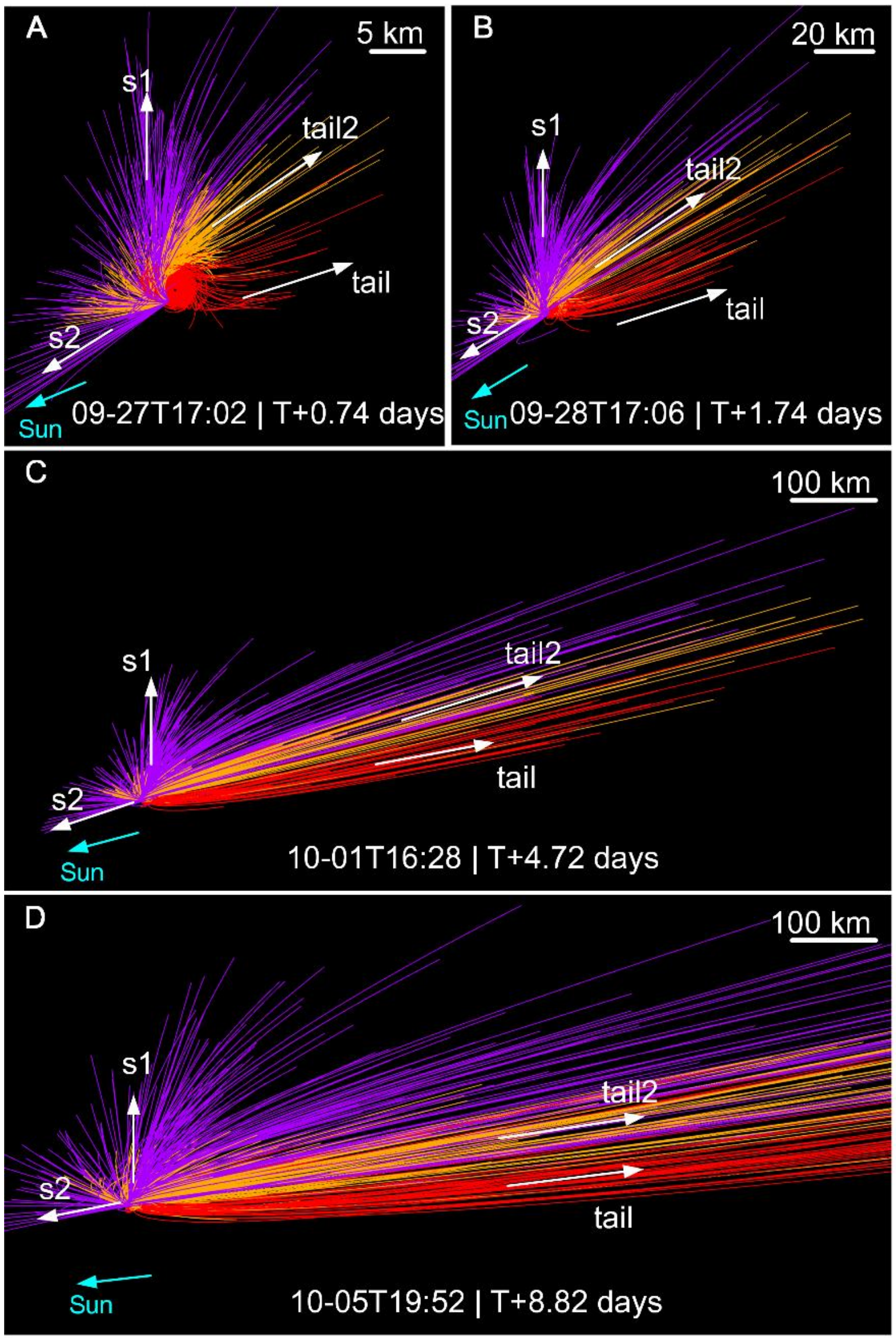

**Fig. S5.**

Trajectories of ejecta particles with radii greater than 100 μm and ejection velocities lower than 0.5 m $s^{-1}$. Purple, orange, and red trajectories correspond to particles with velocities of 0.3–0.5 m $s^{-1}$, 0.25–0.3 m $s^{-1}$, and 0.2–0.25 m $s^{-1}$, respectively. Panel **A** shows that the trajectories of ejecta particles with radii greater than 100 μm but velocities lower than 0.25 m $s^{-1}$ are strongly bent by the system's gravity, causing them to pass over the nightside of Didymos and feed into the tail from its south. In contrast, ejecta with velocities between 0.25–0.3 m $s^{-1}$ experience less gravitational curvature and are mostly injected directly into the tail from the north of Didymos. Ejecta entering the dust tail from the north and south of Didymos progressively developed into the observed bifurcated tail structure (**C** and **D**).

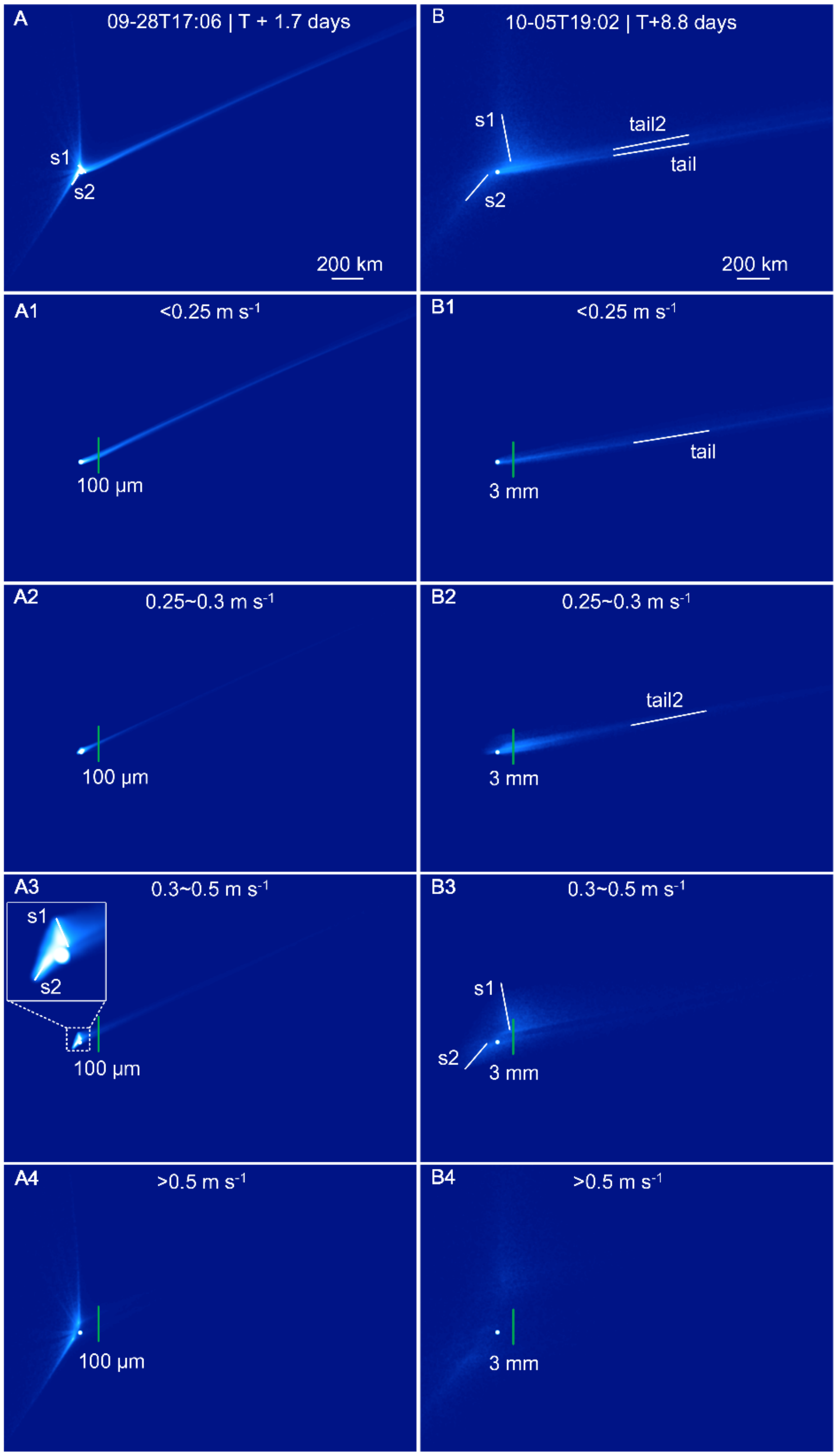

**Fig. S6.**

Evolution of ejecta at different velocities based on the simulations. The first row in each column shows the simulated synthetic image. Rows 2-5 depict the decomposed images corresponding to ejecta with initial velocities less than 0.25 m $s^{-1}$, 0.25~0.3 m $s^{-1}$, 0.3~0.5 m $s^{-1}$, and greater than 0.5 m $s^{-1}$, respectively. The green solid lines in **A1**–**A4** represent the 100 μm size division line, indicating that all escaping primary ejecta particles smaller than 100 μm have been displaced to the right side of this boundary by SRP. The green solid lines in **B1**–**B4** represent the 3 mm size division line, showing that all escaping primary ejecta smaller than 3 mm have been pushed to the right of this line under the effect of SRP. Panels **A1**–**A4** show that micron-sized escaping ejecta are almost entirely injected into the tail under SRP, indicating that gravitational forces have negligible influence on their trajectories during this process. Panels **A3** and **B3** reveal that submillimeter ejecta with ejection velocities between 0.3 and 0.5 m $s^{-1}$ follow trajectories that are bent by the system's gravity before escaping the system. Because the ejecta inherits Dimorphos's orbital velocity, this results in the formation of two ejecta streams (s1 and s2) rotating clockwise from the northern and southern directions. Panel **A4** demonstrates that the trajectories of ejecta with velocities exceeding 0.5 m $s^{-1}$ are hardly perturbed by gravity before escaping the system; these particles form the ejecta cone observed in the early phase, which later evolves into the deflected plume feature (l7, Fig. 1E and F). Panels **B1** and **B2** confirm that the observed double tail consists of submillimeter ejecta with velocities below 0.3 m $s^{-1}$. Specifically, particles with velocities lower than 0.25 m $s^{-1}$ are supplied into the tail from the south, forming the primary tail, while those with velocities between 0.25 and 0.3 m $s^{-1}$ are injected into the tail from the north, giving rise to the secondary tail (Fig. S5).

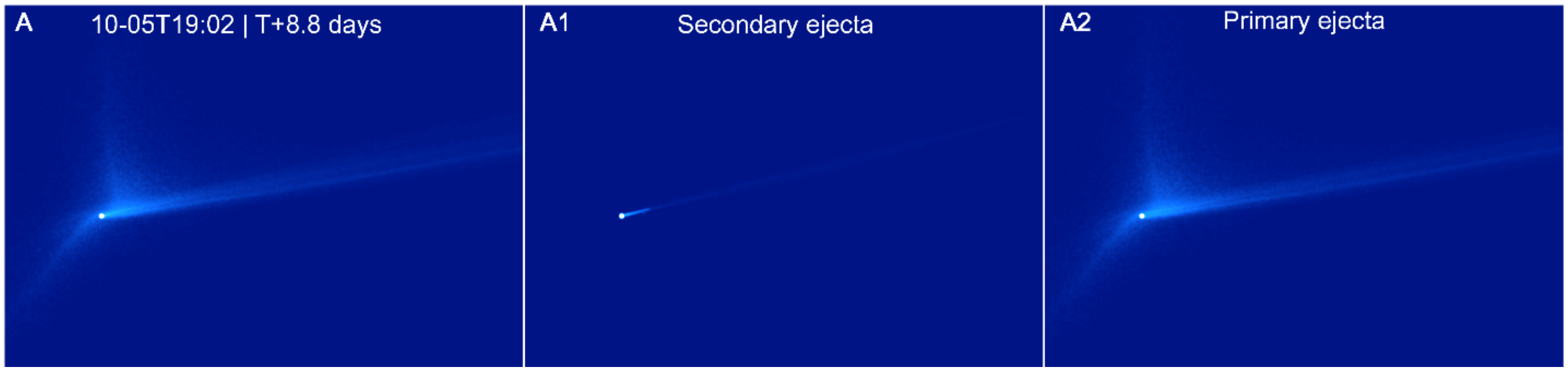

**Fig. S7.**

Decomposed images of ejecta from different sources. All images are displayed with the same logarithmic brightness scale and stretch. (**A**) Original synthetic image, containing both the primary DART ejecta and the ejecta produced by boulder re-impacts. (**A1**) Synthetic image with the tail-like feature generated solely by the secondary ejecta due to boulder re-impacts. (**A2**) Synthetic image with ejecta and tail features produced only by the primary DART ejecta. A comparison between panels **A1** and **A2** demonstrates that the contribution of the secondary ejecta to the tail is negligible compared to that of the primary ejecta.

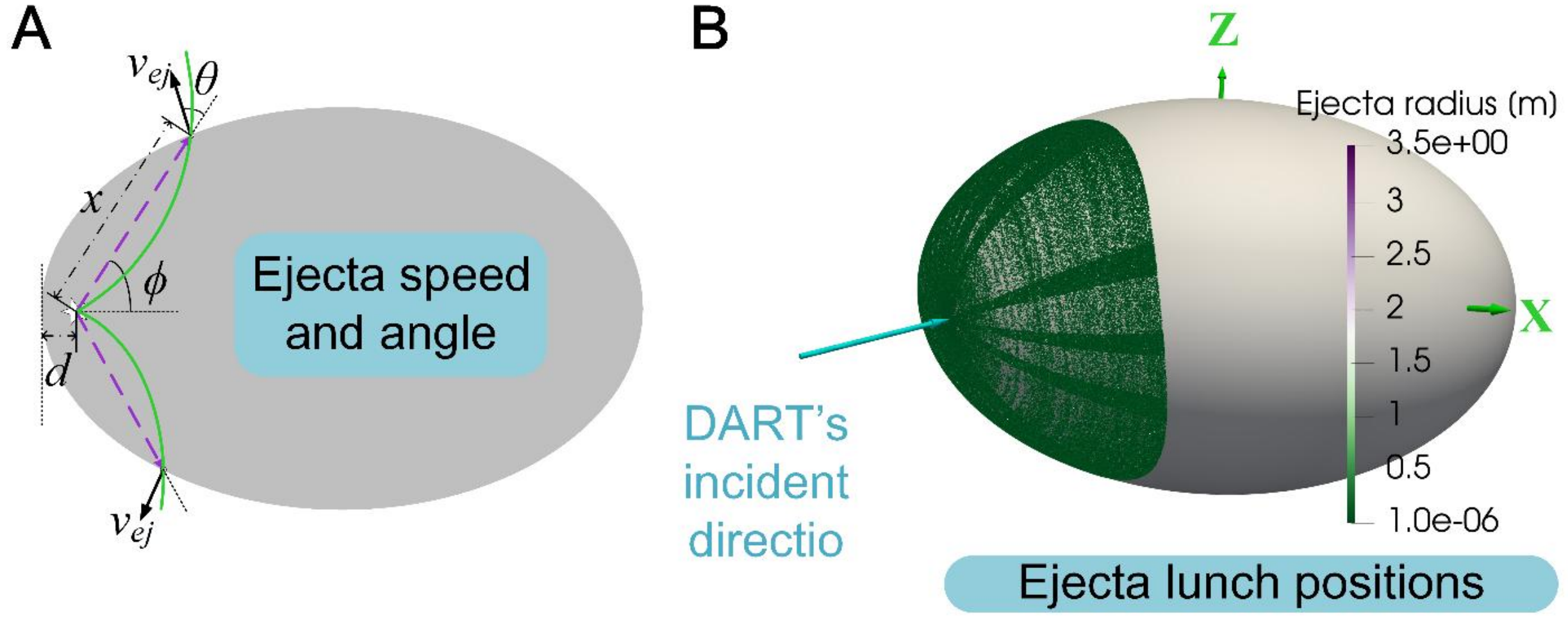

**Fig. S8.**

Generate the ejection positions and velocities of ejecta particles from the surface of Dimorphos. **(A)** Illustrations of streamlines in the Maxwell Z model. A streamline example is shown as a solid green line, where $d$ denotes the distance from the flow center to the impact point, and $x$ is the distance from the ejecta launching location to the flow center. **(B)** Schematic of the distribution of ejecta launching locations on the Dimorphos surface. The color-coded dots on the Dimorphos surface represent the launching locations of ejecta, with colors corresponding to particle radii.

**Table S1.**

Table of iteratively refined parameters vs fixed parameters. As shown in Table S2, since different combinations of scaling law parameters after iterative convergence can all achieve constrained ejecta morphology, we have not provided best fitting values for the scaling law parameters and the Streamtube shape *Z*. Similarly, since each converged solution corresponds to a specific size–frequency distribution of the ejecta, the size–frequency distributions of these converged solutions are provided separately in Table S3.

| **Iteratively refined parameters** | **Range of parameter values** | **initial value** | **Best-fit values** |
|---|---|---|---|
| Scaling parameters $\mu$ | 0.34~0.65 (*30*) | 0.35 | \ |
| Scaling parameters $k$ | 0.3~0.5 (*30*) | random | \ |
| Scaling parameters $C_1$ | 0.18~0.6 (*30*) | random | \ |
| Scaling parameters $H_1$ or $H_2$ | 0.3~0.81 (*30*) | random | \ |
| Streamtube, shape Z ($x$=0) | >2.1 | 2.3 | \ |
| Bulk density of the Dimorphos $\rho_{bulk}$ | 1000~3500 kg $m^{-3}$ (*2*) | random | 1139.10~1689.98 kg $m^{-3}$ |
| dSFD power-law index $q_1$ for 1 um ~ 34 um | -2.0~-5.0 | -2.7 (*9*, *19*) | \ |
| dSFD power-law index $q_2$ for 0.034 mm ~ 2 mm | -2.0~-5.0 | -2.7 (*9*, *19*) | \ |
| dSFD power-law index $q_3$ for 2 mm ~ 1.26 cm | -2.0~-5.0 | -3.7 (*9*, *19*) | \ |
| dSFD power-law index $q_4$ for 1.26 cm ~ 1.0 m | -2.0~-5.0 | -4.2 (*9*, *19*) | \ |
| Weibull distribution for boulders >1.0 m | $\lambda_W$: (0.01 ~0.1), $k_W$: (0.01~0.5) | $\lambda_W$=0.0327, $k_W$=0.39589 (*34*) | \ |
| **Fixed parameters** | **Values** | | |
| Dimorphos dimensions | 177*174*116 m (*1*) | \ | \ |
| Dimorphos volume | 0.00181 $km^3$ (*1*) | \ | \ |
| Density of the Didymos | 2750 kg $m^{-3}$ (*1*) | \ | \ |
| Grain density $\rho$ | 3500 kg $m^{-3}$ (*59*) | \ | \ |
| Cohesion *C* | 0, 5, 50, 250 Pa (*16*) | \ | \ |
| Flow center depth *d* | 0~1 times DART radius with 0.1 intervals (*44*) | \ | 0.5 |
| Streamtube, shape Z ($x$=$x_{max}$) | 2.5~3.3 with 0.1 intervals (*18*) | \ | 3.0 |

**Table S2.**

Initial input parameters and converged parameters for each simulation.

| Identifier | Parameter set | Bulk density (kg m$^{-3}$) | Cohesion strength (Pa) | $\mu$ | $k$ | $C_1$ | $H_1$ or $H_2$ | $Z$ ($x$ = 0) | $\beta_D$ | $\beta_{sys}$ |
|---|---|---|---|---|---|---|---|---|---|---|
| 1 | Initial parameters | 2424.280 | 0 | 0.350000 | 0.467624 | 0.562782 | 0.466454 | 2.3 | 1.841083 | 1.804830 |
| | Converged parameters | 1648.612 | 0 | 0.385761 | 0.467624 | 0.583600 | 0.441826 | 2.207910 | 2.390306 | 2.365244 |
| 2 | Initial parameters | 2241.176 | 0 | 0.350000 | 0.443729 | 0.528232 | 0.396795 | 2.3 | 1.759983 | 1.730944 |
| | Converged parameters | 1588.346 | 0 | 0.395170 | 0.443729 | 0.559400 | 0.371822 | 2.200000 | 2.355459 | 2.257925 |
| 3 | Initial parameters | 1356.129 | 0 | 0.350000 | 0.378052 | 0.476129 | 0.391172 | 2.3 | 1.892719 | 1.849274 |
| | Converged parameters | 1409.362 | 0 | 0.400104 | 0.378052 | 0.516793 | 0.364219 | 2.221000 | 2.069853 | 1.986455 |
| 4 | Initial parameters | 2976.946 | 0 | 0.350000 | 0.419625 | 0.571342 | 0.387737 | 2.3 | 1.866831 | 1.845882 |
| | Converged parameters | 1601.778 | 0 | 0.394722 | 0.419625 | 0.596305 | 0.361114 | 2.200000 | 2.360862 | 2.285678 |
| 5 | Initial parameters | 2111.900 | 0 | 0.350000 | 0.475837 | 0.336196 | 0.363910 | 2.3 | 1.625175 | 1.600515 |
| | Converged parameters | 1420.942 | 0 | 0.400526 | 0.475837 | 0.380959 | 0.333646 | 2.200000 | 2.055632 | 1.978694 |
| 6 | Initial parameters | 1534.125 | 5 | 0.350000 | 0.469843 | 0.273929 | 0.346094 | 2.3 | 1.365998 | 1.350118 |
| | Converged parameters | 1406.702 | 5 | 0.428905 | 0.469843 | 0.325071 | 0.315431 | 2.200000 | 2.097089 | 2.008987 |
| 7 | Initial parameters | 3492.169 | 5 | 0.350000 | 0.434560 | 0.581125 | 0.594276 | 2.3 | 1.617203 | 1.583706 |
| | Converged parameters | 1450.402 | 5 | 0.393229 | 0.434560 | 0.600000 | 0.568192 | 2.200000 | 2.209737 | 2.096439 |
| 8 | Initial parameters | 2979.613 | 5 | 0.350000 | 0.307479 | 0.554818 | 0.541536 | 2.3 | 1.437479 | 1.422704 |
| | Converged parameters | 1501.651 | 5 | 0.407516 | 0.307479 | 0.579656 | 0.510776 | 2.216869 | 2.167850 | 2.082399 |
| 9 | Initial parameters | 2391.405 | 5 | 0.350000 | 0.497678 | 0.377132 | 0.368449 | 2.3 | 1.600037 | 1.606657 |
| | Converged parameters | 1474.030 | 5 | 0.424951 | 0.497678 | 0.334418 | 0.336614 | 2.200000 | 2.244664 | 2.114577 |
| 10 | Initial parameters | 1692.872 | 5 | 0.350000 | 0.315364 | 0.190902 | 0.416690 | 2.3 | 1.165413 | 1.156601 |
| | Converged parameters | 1139.105 | 5 | 0.427454 | 0.315364 | 0.274350 | 0.383246 | 2.212306 | 1.701267 | 1.632087 |

| Identifier | Parameter set | Bulk density (kg m$^{-3}$) | Cohesion strength (Pa) | $\mu$ | $k$ | $C_1$ | $H_1$ or $H_2$ | $Z$ ($x$ = 0) | $\beta_D$ | $\beta_{sys}$ |
|---|---|---|---|---|---|---|---|---|---|---|
| 11 | Initial parameters | 3382.176 | 50 | 0.350000 | 0.418185 | 0.301347 | 0.658002 | 2.3 | 1.302986 | 1.296396 |
| | Converged parameters | 1357.562 | 50 | 0.416751 | 0.418185 | 0.362829 | 0.626040 | 2.200000 | 2.076022 | 1.952908 |
| 12 | Initial parameters | 2757.103 | 50 | 0.350000 | 0.317776 | 0.553785 | 0.774464 | 2.3 | 1.439539 | 1.421143 |
| | Converged parameters | 1375.390 | 50 | 0.407511 | 0.317776 | 0.598143 | 0.745385 | 2.200000 | 2.117971 | 1.986045 |
| 13 | Initial parameters | 2158.450 | 50 | 0.350000 | 0.462643 | 0.510312 | 0.703019 | 2.3 | 1.677090 | 1.649807 |
| | Converged parameters | 1566.899 | 50 | 0.404551 | 0.462643 | 0.542048 | 0.676210 | 2.200000 | 2.365375 | 2.255313 |
| 14 | Initial parameters | 2774.771 | 50 | 0.350000 | 0.495711 | 0.404878 | 0.799661 | 2.3 | 1.502744 | 1.477376 |
| | Converged parameters | 1520.124 | 50 | 0.400463 | 0.495711 | 0.440042 | 0.771850 | 2.200000 | 2.193275 | 2.144509 |
| 15 | Initial parameters | 3221.875 | 50 | 0.350000 | 0.379084 | 0.220988 | 0.740943 | 2.3 | 1.220169 | 1.212613 |
| | Converged parameters | 1181.253 | 50 | 0.418471 | 0.379084 | 0.289744 | 0.708264 | 2.226776 | 1.775773 | 1.637881 |
| 16 | Initial parameters | 2662.105 | 50 | 0.350000 | 0.361999 | 0.286750 | 0.670915 | 2.3 | 1.299434 | 1.288365 |
| | Converged parameters | 1295.288 | 50 | 0.419465 | 0.361999 | 0.348218 | 0.638325 | 2.207879 | 1.862465 | 1.811956 |
| 17 | Initial parameters | 1416.406 | 50 | 0.350000 | 0.423057 | 0.242318 | 0.457431 | 2.3 | 1.295539 | 1.291815 |
| | Converged parameters | 1356.310 | 50 | 0.450368 | 0.423057 | 0.268494 | 0.424642 | 2.200000 | 2.048303 | 1.990003 |
| 18 | Initial parameters | 1141.606 | 50 | 0.350000 | 0.382162 | 0.551704 | 0.635289 | 2.3 | 1.709187 | 1.701166 |
| | Converged parameters | 1547.768 | 50 | 0.420190 | 0.382162 | 0.482047 | 0.607813 | 2.201703 | 2.295421 | 2.2100480 |
| 19 | Initial parameters | 2781.059 | 250 | 0.350000 | 0.453513 | 0.377365 | 0.674873 | 2.3 | 1.474228 | 1.465739 |
| | Converged parameters | 1434.699 | 250 | 0.442969 | 0.453513 | 0.268756 | 0.644727 | 2.200000 | 2.136842 | 2.048600 |
| 20 | Initial parameters | 2908.342 | 250 | 0.350000 | 0.415082 | 0.449065 | 0.714723 | 2.3 | 1.532464 | 1.527584 |
| | Converged parameters | 1389.943 | 250 | 0.439651 | 0.415082 | 0.322361 | 0.685023 | 2.200000 | 2.096143 | 2.037979 |
| 21 | Initial parameters | 2111.900 | 0 | 0.350000 | 0.475837 | 0.336196 | 0.363910 | 2.3 | 1.625175 | 1.600515 |
| | Converged parameters | 1420.942 | 0 | 0.400526 | 0.475837 | 0.380959 | 0.333646 | 2.200000 | 2.055632 | 1.978694 |

| Identifier | Parameter set | Bulk density (kg m$^{-3}$) | Cohesion strength (Pa) | $\mu$ | $k$ | $C_1$ | $H_1$ or $H_2$ | $Z$ ($x$ = 0) | $\beta_D$ | $\beta_{sys}$ |
|---|---|---|---|---|---|---|---|---|---|---|
| 22 | Initial parameters | 2225.279 | 5 | 0.350000 | 0.489061 | 0.531300 | 0.681974 | 2.3 | 1.773227 | 1.743603 |
| | Converged parameters | 1619.009 | 5 | 0.387463 | 0.489061 | 0.551140 | 0.657657 | 2.225915 | 2.446707 | 2.363226 |
| 23 | Initial parameters | 1817.686 | 250 | 0.350000 | 0.462578 | 0.490405 | 0.412257 | 2.3 | 1.578140 | 1.578887 |
| | Converged parameters | 1431.161 | 250 | 0.458424 | 0.462578 | 0.182787 | 0.381356 | 2.200000 | 1.967315 | 1.861017 |
| 24 | Initial parameters | 2975.569 | 0 | 0.350000 | 0.470195 | 0.354184 | 0.426026 | 2.3 | 1.475751 | 1.452908 |
| | Converged parameters | 1445.361 | 0 | 0.397192 | 0.470195 | 0.393634 | 0.394991 | 2.212310 | 2.098569 | 1.996748 |
| 25 | Initial parameters | 3438.935 | 50 | 0.350000 | 0.389781 | 0.365559 | 0.498083 | 2.3 | 1.387393 | 1.367266 |
| | Converged parameters | 1329.342 | 50 | 0.444951 | 0.389781 | 0.300038 | 0.466604 | 2.203638 | 2.035462 | 1.950274 |
| 26 | Initial parameters | 2958.881 | 50 | 0.350000 | 0.434851 | 0.479996 | 0.500268 | 2.3 | 1.529140 | 1.520959 |
| | Converged parameters | 1351.492 | 50 | 0.441291 | 0.434851 | 0.270443 | 0.470579 | 2.200000 | 2.052761 | 1.973747 |
| 27 | Initial parameters | 1813.004 | 50 | 0.350000 | 0.305731 | 0.539094 | 0.570456 | 2.3 | 1.497769 | 1.486511 |
| | Converged parameters | 1437.591 | 50 | 0.438319 | 0.305731 | 0.407193 | 0.540520 | 2.220718 | 2.082155 | 2.015212 |
| 28 | Initial parameters | 2667.879 | 50 | 0.350000 | 0.399015 | 0.243851 | 0.554880 | 2.3 | 1.198408 | 1.186660 |
| | Converged parameters | 1329.406 | 50 | 0.432654 | 0.399015 | 0.305456 | 0.522005 | 2.200000 | 2.010857 | 1.918437 |
| 29 | Initial parameters | 2779.664 | 250 | 0.350000 | 0.485268 | 0.215996 | 0.312850 | 2.3 | 1.338919 | 1.344337 |
| | Converged parameters | 1281.467 | 250 | 0.445826 | 0.485268 | 0.180000 | 0.676522 | 2.200000 | 1.983841 | 1.889213 |
| 30 | Initial parameters | 1076.456 | 50 | 0.350000 | 0.468039 | 0.466511 | 0.671396 | 2.3 | 1.791428 | 1.768563 |
| | Converged parameters | 1633.543 | 50 | 0.411225 | 0.468039 | 0.481116 | 0.644660 | 2.200000 | 2.390621 | 2.294314 |
| 31 | Initial parameters | 3474.459 | 50 | 0.350000 | 0.499799 | 0.555052 | 0.781253 | 2.3 | 1.548945 | 1.542116 |
| | Converged parameters | 1689.977 | 50 | 0.398129 | 0.499799 | 0.571423 | 0.754233 | 2.200000 | 2.482553 | 2.420913 |
| 32 | Initial parameters | 3383.177 | 0 | 0.350000 | 0.376791 | 0.376688 | 0.461400 | 2.3 | 1.448771 | 1.431710 |
| | Converged parameters | 1290.382 | 0 | 0.395098 | 0.376791 | 0.412529 | 0.431312 | 2.221139 | 1.911744 | 1.817798 |

**Table S3.**

Converged ejecta size distributions and truncation radii for the simulations listed in Table S2. The four power-law indices describe dN/dr proportional to r raised to q. Their radius intervals are $q_1$: [$x_4$, $x_3$], $q_2$: [$x_3$, $x_2$], $q_3$: [$x_2$, $x_1$], $q_4$: [$x_1$, $x_0$]. The Weibull segment occupies [$x_0$, $x_d$], with shape $k_w$ and scale $\lambda_w$. All radii and the Weibull scale are in metres. $x_4$ and $x_d$ are the lower and upper truncation radii. $x_3$, $x_2$, $x_1$ and $x_0$ are transition radii.

| Identifier | $q_1$ ($x_4$–$x_3$) | $q_2$ ($x_3$–$x_2$) | $q_3$ ($x_2$–$x_1$) | $q_4$ ($x_1$–$x_0$) | $k_W$ | $\lambda_W$ (m) | $x_4$ (m) | $x_3$ (m) | $x_2$ (m) | $x_1$ (m) | $x_0$ (m) | $x_d$ (m) |
|---|---|---|---|---|---|---|---|---|---|---|---|---|
| 1 | -3.1902 | -2.2636 | -3.1705 | -4.5436 | 0.2217 | 0.0222 | 1e-06 | 3.366e-05 | 0.002063 | 0.01217 | 1.096 | 3.5 |
| 2 | -3.1917 | -2.2779 | -3.1720 | -4.5360 | 0.2217 | 0.0222 | 1e-06 | 3.386e-05 | 0.002049 | 0.01227 | 1.092 | 3.5 |
| 3 | -3.1900 | -2.2700 | -3.1500 | -4.5000 | 0.2214 | 0.02222 | 1e-06 | 3.4e-05 | 0.0021 | 0.0126 | 1.07 | 3.5 |
| 4 | -3.1950 | -2.2828 | -3.1685 | -4.5449 | 0.2217 | 0.0222 | 1e-06 | 3.406e-05 | 0.002067 | 0.01219 | 1.097 | 3.5 |
| 5 | -3.1942 | -2.2781 | -3.1720 | -4.5457 | 0.2217 | 0.0222 | 1e-06 | 3.393e-05 | 0.002057 | 0.01214 | 1.098 | 3.5 |
| 6 | -3.1987 | -2.3042 | -3.1752 | -4.4939 | 0.2210 | 0.02224 | 1e-06 | 3.445e-05 | 0.002043 | 0.01231 | 1.04 | 3.5 |
| 7 | -3.1900 | -2.2700 | -3.1500 | -4.5000 | 0.2214 | 0.02222 | 1e-06 | 3.4e-05 | 0.0021 | 0.0126 | 1.07 | 3.5 |
| 8 | -3.1981 | -2.2797 | -3.1646 | -4.5146 | 0.2221 | 0.02218 | 1e-06 | 3.451e-05 | 0.002061 | 0.01232 | 1.086 | 3.5 |
| 9 | -3.1900 | -2.2700 | -3.1500 | -4.5000 | 0.2214 | 0.02222 | 1e-06 | 3.4e-05 | 0.0021 | 0.0126 | 1.07 | 3.5 |
| 10 | -3.1394 | -2.3465 | -3.2128 | -4.5084 | 0.2198 | 0.02445 | 1e-06 | 3.493e-05 | 0.001929 | 0.01115 | 0.9493 | 3.5 |
| 11 | -3.1995 | -2.2815 | -3.1707 | -4.5038 | 0.2240 | 0.02228 | 1e-06 | 3.429e-05 | 0.002057 | 0.01215 | 1.073 | 3.5 |
| 12 | -3.1647 | -2.3083 | -3.1814 | -4.5042 | 0.2206 | 0.02331 | 1e-06 | 3.446e-05 | 0.002013 | 0.01185 | 1.008 | 3.5 |
| 13 | -3.1647 | -2.3083 | -3.1814 | -4.5042 | 0.2206 | 0.02331 | 1e-06 | 3.446e-05 | 0.002013 | 0.01185 | 1.008 | 3.5 |
| 14 | -3.2055 | -2.2895 | -3.1792 | -4.5877 | 0.2182 | 0.02245 | 1e-06 | 3.466e-05 | 0.002023 | 0.01205 | 1.038 | 3.5 |
| 15 | -3.1900 | -2.2700 | -3.1500 | -4.5000 | 0.2214 | 0.02222 | 1e-06 | 3.4e-05 | 0.0021 | 0.01053 | 1.07 | 3.5 |
| 16 | -3.1996 | -2.3072 | -3.1958 | -4.5026 | 0.2208 | 0.02226 | 1e-06 | 3.441e-05 | 0.002016 | 0.01082 | 1.023 | 3.5 |
| 17 | -3.2053 | -2.3362 | -3.1998 | -4.4752 | 0.2203 | 0.02229 | 1e-06 | 3.485e-05 | 0.001998 | 0.01197 | 0.9906 | 3.5 |
| 18 | -3.1976 | -2.3045 | -3.1797 | -4.4954 | 0.2210 | 0.02225 | 1e-06 | 3.442e-05 | 0.002037 | 0.01223 | 1.037 | 3.5 |
| 19 | -3.1900 | -2.2700 | -3.1500 | -4.5000 | 0.2214 | 0.02222 | 1e-06 | 3.4e-05 | 0.0021 | 0.0126 | 1.07 | 3.5 |
| 20 | -3.1900 | -2.2700 | -3.1500 | -4.5000 | 0.2214 | 0.02222 | 1e-06 | 3.4e-05 | 0.0021 | 0.0126 | 1.07 | 3.5 |
| 21 | -3.1942 | -2.2781 | -3.1720 | -4.5457 | 0.2217 | 0.0222 | 1e-06 | 3.393e-05 | 0.002057 | 0.01214 | 1.098 | 3.5 |
| 22 | -3.1900 | -2.2700 | -3.1500 | -4.6170 | 0.2214 | 0.02222 | 1e-06 | 3.4e-05 | 0.0021 | 0.0126 | 1.07 | 3.5 |
| 23 | -3.1900 | -2.2700 | -3.1500 | -4.5000 | 0.2214 | 0.02222 | 1e-06 | 3.4e-05 | 0.0021 | 0.0126 | 1.07 | 3.5 |
| 24 | -3.2038 | -2.2713 | -3.1957 | -4.5775 | 0.2265 | 0.02265 | 1e-06 | 3.353e-05 | 0.002031 | 0.01147 | 1.068 | 3.5 |
| 25 | -3.1973 | -2.2991 | -3.1689 | -4.4926 | 0.2210 | 0.02224 | 1e-06 | 3.439e-05 | 0.002055 | 0.01243 | 1.044 | 3.5 |
| 26 | -3.1900 | -2.2700 | -3.1500 | -4.5000 | 0.2214 | 0.02222 | 1e-06 | 3.4e-05 | 0.0021 | 0.0126 | 1.07 | 3.5 |
| 27 | -3.1900 | -2.2700 | -3.1500 | -4.5000 | 0.2214 | 0.02222 | 1e-06 | 3.4e-05 | 0.0021 | 0.0126 | 1.07 | 3.5 |
| 28 | -3.2034 | -2.3264 | -3.2106 | -4.4713 | 0.2202 | 0.02229 | 1e-06 | 3.468e-05 | 0.001987 | 0.01169 | 0.9824 | 3.5 |
| 29 | -3.1900 | -2.2700 | -3.1500 | -4.5000 | 0.2214 | 0.02222 | 1e-06 | 3.4e-05 | 0.0021 | 0.0126 | 1.07 | 3.5 |
| 30 | -3.1900 | -2.2700 | -3.1500 | -4.5000 | 0.2214 | 0.02222 | 1e-06 | 3.4e-05 | 0.0021 | 0.0126 | 1.07 | 3.5 |
| 31 | -3.1900 | -2.2700 | -3.1500 | -4.5000 | 0.2214 | 0.02222 | 1e-06 | 3.4e-05 | 0.0021 | 0.0126 | 1.07 | 3.5 |
| 32 | -3.2081 | -2.2350 | -3.1997 | -4.6776 | 0.2206 | 0.02226 | 1e-06 | 3.247e-05 | 0.002035 | 0.01143 | 1.006 | 3.5 |

**Table S4.**

Weights corresponding to the data at each observation time.

| Observation time | T+3.5 h | T+5.0 h | T+8.2 h | T+0.7 days | T+1.7 days | T+2.1 days | T+3.7 days | T+4.7 days | T+5.7 days | T+8.8 days | T+11.9 days | T+14.9 days | T+18.5 days |
|---|---|---|---|---|---|---|---|---|---|---|---|---|---|
| Weight | 0.0083 | 0.0083 | 0.0083 | 0.331 | 0.331 | 0.017 | 0.0167 | 0.099 | 0.099 | 0.0331 | 0.0331 | 0.0083 | 0.0083 |

## Supplementary Materials

Materials and Methods

Figs. S1 to S8

Tables S1 to S4